\documentclass[12pt,a4paper]{report}

\usepackage[top=0.70in, bottom=0.70in, left=0.8in,right=0.80in]{geometry} 
\usepackage[pdftex]{graphicx} 
\usepackage[final]{pdfpages} 
\usepackage{graphicx}
\usepackage{url}
\usepackage{subfigure}
\usepackage{makecell}
\usepackage{listings}
\usepackage{paralist}
\let\subparagraph\paragraph
\usepackage{amsfonts, amsmath, amssymb}
\usepackage{algorithm}
\usepackage{algorithmic}
\usepackage{epstopdf}
\usepackage{float}
\usepackage{enumerate}
\usepackage{color}
\usepackage{multirow}
\usepackage{verbatim}
\usepackage{stfloats}
\usepackage[framemethod=tikz]{mdframed}
\usepackage{lipsum}
\usepackage{theorem}
\usepackage{cite}
\usepackage{longtable}
\usepackage[figuresright]{rotating}
\usepackage{geometry}
\usepackage[T1]{fontenc}
\usepackage[ansinew]{inputenc}
\newtheorem{mydef}{Definition}

\theorembodyfont{\bfseries\upshape}

\newcommand{\gt}[1]{\texttt{#1}}

\newcommand{\fp}{\textsc{$f_p$}}
\newcommand{\av}{\textsc{AV}}

\newcommand{\ccsl}{\textsc{Ccsl}}

\newcommand{\uppaal}{\textsc{Uppaal}}
\newcommand{\smc}{\textsc{Uppaal-SMC}}

\newcommand{\simu}{\textsc{Simulink}}
\newcommand{\staf}{\textsc{Stateflow}}

\newcommand{\ed}{\textsc{East-adl}}
\newcommand{\auto}{\textsc{AUTOSAR}}
\newcommand{\ft}{\textsc{$f_t$}}

\newcommand{\marte}{\textsc{MARTE}}
\newcommand{\prccsl}{PrCCSL}
\newcommand{\tdl}{\textsc{Tadl2}}

\usepackage[font=small,labelfont=bf]{caption}

\usepackage{listings}
\usepackage{color}

\definecolor{dkgreen}{rgb}{0,0.6,0}
\definecolor{gray}{rgb}{0.5,0.5,0.5}
\definecolor{mauve}{rgb}{0.58,0,0.82}

\usepackage{fancyhdr}
\fancypagestyle{plain}{%
\fancyfoot[L]{\emph{}} 
\fancyfoot[R]{\thepage}

}

\usepackage{pgf}
\usepackage{pgfpages}

\pgfpagesdeclarelayout{boxed}
{
  \edef\pgfpageoptionborder{0pt}
}
{
  \pgfpagesphysicalpageoptions
  {%
    logical pages=1,%
  }
  \pgfpageslogicalpageoptions{1}
  {
    border code=\pgfsetlinewidth{2pt}\pgfstroke,%
    border shrink=\pgfpageoptionborder,%
    resized width=.95\pgfphysicalwidth,%
    resized height=.95\pgfphysicalheight,%
    center=\pgfpoint{.5\pgfphysicalwidth}{.5\pgfphysicalheight}%
  }%
}
\begin{document}
\renewcommand\bibname{References}
\lhead{ }



\newpage
\begin{center}
\thispagestyle{empty}
\setlength{\voffset}{2in}
\vspace{2cm}
\LARGE{\textbf{Probabilistic Analysis of Weakly-Hard Real-Time Systems}}\\
\vspace{2cm}
{\Large{\textbf{Eun-Young Kang$^{12}$ and Dongrui Mu$^{2}$ and Li Huang$^{2}$}}}\\
\vspace{0.3cm}
\large{\textbf{$^1$University of Namur, Belgium}}\\
\vspace{0.3cm}
\large{\textbf{$^2$School of Data \& Computer Science, Sun Yat-Sen University, China}}\\
\vspace{0.3cm}
\Large{{eykang@fundp.ac.be}}\\
\Large{{\{mudr, huangl223\}@mail2.sysu.edu.cn}}\\
\vspace{3cm}
\newpage

\end{center}
\newpage



\begin{center}
\thispagestyle{empty}
\vspace{2cm}
\LARGE{\textbf{ABSTRACT}}\\[1.0cm]
\end{center}
\thispagestyle{empty}
\large{\paragraph{}
Modeling and analysis of non-functional properties, such as timing constraints, is crucial in automotive real-time embedded systems. \ed\ is a domain specific architectural language dedicated to safety-critical automotive embedded system design. We have previously specified \ed\ timing constraints in Clock Constraint Specification Language (\ccsl) and proved the correctness of specification by mapping the semantics of the constraints into \uppaal\ models amenable to model checking. In most cases, a bounded number of violations of timing constraints in automotive systems would not lead to system failures when the results of the violations are negligible, called Weakly-Hard (WH). Previous work is extended in this paper by including support for probabilistic analysis of timing constraints in the context of WH: Probabilistic extension of \ccsl, called Pr\ccsl, is defined and the \ed\ timing constraints with stochastic properties are specified in Pr\ccsl. The semantics of the extended constraints in Pr\ccsl\ is translated into \smc\ models for formal verification. Furthermore, a set of mapping rules is proposed to facilitate guarantee of translation. Our approach is demonstrated on an autonomous traffic sign recognition vehicle case study.}

\newpage

\pagenumbering{roman} 

\pagestyle{empty}
\addtocontents{toc}{\protect\thispagestyle{empty}}
\tableofcontents 

\addtocontents{lof}{\protect\thispagestyle{empty}}
\listoffigures 
\cleardoublepage

\pagestyle{fancy}

\newpage
\pagenumbering{arabic} 

\chapter{Introduction}
\label{sec:intro}

Model-driven development is rigorously applied in automotive systems in which the software controllers interact with physical environments. The continuous time behaviors (evolved with various energy rates) of those systems often rely on complex dynamics as well as on stochastic behaviors. Formal verification and validation (V\&V) technologies are indispensable and highly recommended for development of safe and reliable automotive systems \cite{iso26262,iec61508}. Conventional V\&V, i.e., testing and model checking have limitations in terms of assessing the reliability of hybrid systems due to both the stochastic and non-linear dynamical features. To ensure the reliability of safety critical hybrid dynamic systems, \emph{statistical model checking (SMC)} techniques have been proposed \cite{smc-challenge,smc-david-12,david2015uppaal}. These techniques for fully stochastic models validate probabilistic performance properties of given deterministic (or stochastic) controllers in given stochastic environments.

Conventional formal analysis of timing models addresses worst case
designs, typically used for hard deadlines in safety critical systems,
however, there is great incentive to include ``less-than-worst-case''
designs to improve efficiency but without affecting the quality
of timing analysis in the systems. The challenge is the definition of
suitable model semantics that provide reliable predictions of \emph{system
timing}, given the timing of individual components and their
compositions. While the standard worst case models are well understood
in this respect, the behavior and the expressiveness of
``less-than-worst-case'' models is far less investigated. In most cases,
a bounded number of violations of timing constraints in
systems would not lead to system failures when the results of the
violations are negligible, called Weakly-Hard (WH)
\cite{Nicolau1988Specification, Bernat2001Weakly}. In this paper, we
propose a formal probabilistic modeling and analysis technique by
extending the known concept of WH constraints to what is called
``typical'' worst case model and analysis.

\ed\ (Electronics Architecture and Software Technology - Architecture
Description Language) \cite{EAST-ADL,maenad}, aligned with \auto\
(Automotive Open System Architecture) standard \cite{autosar}, is a
concrete example of the MBD approach for the architectural modeling
of safety-critical automotive embedded systems. A system in \ed\ is described by {\gt{Functional Architectures (FA)}} at different abstraction levels. The {\gt{FA}} are composed of a number of interconnected \emph{functionprototypes} (\fp), and the \fp s have ports and connectors for communication. \ed\ relies on external tools for the analysis of specifications related to requirements. For example, behavioral description in \ed\ is captured in external tools, i.e., \simu/\staf \cite{slsf}.
The latest release of \ed\ has adopted the time model proposed in the Timing Augmented Description Language (\tdl) \cite{TADL2}. \tdl\ expresses and
composes the basic timing constraints, i.e., repetition rates,
End-to-End delays, and synchronization constraints. The time model
of \tdl\ specializes the time model of \marte, the UML profile for
Modeling and Analysis of Real-Time and Embedded systems
\cite{MARTE}. \marte\ provides \ccsl, a time model and a Clock
Constraint Specification Language, that supports specification of
both logical and dense timing constraints for \marte\ models, as well
as functional causality constraints \cite{ccsl_correctness}.

We have previously specified non-functional properties (timing and energy constraints) of automotive systems specified in \ed\ and \marte/\ccsl, and proved the correctness of specification by mapping the semantics of the constraints into \uppaal\ models for model checking \cite{ksac14}.
Previous work is extended in this paper by including support
for probabilistic analysis of timing constraints of automotive systems in the context WH: \begin{inparaenum} \item Probabilistic extension of \ccsl, called Pr\ccsl, is
defined and the \ed/\tdl\ timing constraints with stochastic properties
are specified in Pr\ccsl; \item The semantics of the extended constraints in Pr\ccsl\ is translated into verifiable \smc\ \cite{smc} models for formal verification; \item A set of mapping rules is
proposed to facilitate guarantee of translation. \end{inparaenum} Our approach is demonstrated on an autonomous traffic sign recognition vehicle (AV) case study.

The paper is organized as follows: Chapter \ref{sec:preliminary}
presents an overview of \ccsl\ and \smc. The AV is introduced as a  running example in Chapter \ref{sec:case-study}. Chapter \ref{sec:def_ccsl} presents the formal definition of Pr\ccsl. The timing constraints that are applied on top of AV are specified using \ccsl\ in Chapter \ref{sec:specification}. Chapter \ref{sec:model} describes a set of translation patterns from \ccsl/Pr\ccsl\ to \smc\  models and how our approaches provide support for formal analysis at the design level. The behaviours of AV system and the stochastic behaviours of the environments are represented as a network of Stochastic Timed Automata presented in Chapter \ref{sec:beha}.
The applicability of our method is demonstrated by performing verification on the AV case study in Chapter \ref{sec:experiment}. Chapter \ref{sec:r-work} and Chapter \ref{sec:conclusion} present related work and the conclusion.

\chapter{preliminary}
\label{sec:preliminary}

In our framework, we consider a subset of \ccsl\ and its extension with stochastic properties that is sufficient to specify \ed\ timing constraints in the context of WH. Formal Modeling and V\&V of the \ed\ timing constraints specified in \ccsl\ are performed using \smc.

\vspace{0.05in}
\noindent \textbf{Clock Constraint Specification Language} (\ccsl) \cite{ccsl_correctness, andre2009syntax} is a UML profile for modeling and analysis of real-time systems (\marte) \cite{Andr2008Clock,abs-ccsl}. In \ccsl,  a clock represents a sequence of (possibly infinite) instants.
An event is a clock and the occurrences of an event correspond to a set of ticks of the clock.
\ccsl\ provides a set of clock constraints that specifies evolution of clocks' ticks. The physical time is represented by a dense clock with a base unit. A dense clock can be discretized into a discrete/logical clock. $idealClock$ is a predefined dense clock whose unit is second. We define a universal clock $ms$ based on $idealClock$: $ms$ = $idealClock$ {\gt{discretizedBy}} 0.001. $ms$ representing a periodic clock that ticks every 1 millisecond in this paper. A step is a tick of the universal clock. Hence the length of one step is 1 millisecond.

\ccsl\ provides two types of clock constraints, \emph{relation} and \emph{expression}: A \emph{relation} limits the occurrences among different events/clocks. Let $C$ be a set of clocks, $c1, c2 \in C$, {\gt{coincidence}} relation ($c1$ $\equiv$ $c2$) specifies that two clocks must tick simultaneously.  {\gt {Precedence}} relation ($c1 \prec c2$) delimits that $c1$ runs faster than $c2$, i.e., $\forall k \in \mathbb{N^{+}}$, where $\mathbb{N^{+}}$ is the set of positive natural numbers, the $k^{th}$ tick of $c1$ must occur prior to the $k^{th}$ tick of $c2$. {\gt{Causality}} relation ($c1 \preceq c2$) represents a relaxed version of {\gt {precedence}}, allowing the two clocks to tick at the same time. {\gt{Subclock}} ($c1$ $\subseteq$ $c2$) indicates the relation between two clocks, \emph{superclock} ($c1$) and \emph{subclock} ($c2$), s.t. each tick of the subclock must correspond to a tick of its superclock at the same step. {\gt{Exclusion}} ($c1$ \# $c2$) prevents the instants of two clocks from being coincident. An \emph{expression} derives new clocks from the already defined clocks:  {\gt{periodicOn}} builds a new clock based on a \emph{base} clock and a \emph{period} parameter, s.t., the instants of the new clock are separated by a number of instants of the \emph{base} clock. The number is given as \emph{period}. {\gt{DelayFor}} results in a clock by delaying the \emph{base} clock for a given number of ticks of a \emph{reference} clock. {\gt{Infimum}}, denoted {\gt{inf}}, is defined as the slowest clock that is faster than both $c1$ and $c2$.  {\gt{Supremum}}, denoted {\gt{sup}}, is defined as the fastest clock that is slower than $c1$ and $c2$.

\vspace{0.05in}
\noindent \textbf{UPPAAL-SMC} performs the probabilistic analysis of properties by monitoring simulations of complex hybrid systems in a given stochastic environment and using results from the statistics to determine whether the system satisfies the property with some degree of confidence. Its clocks evolve with various rates, which are specified with \emph{ordinary differential equations} (ODE). \smc\ provides a number of queries related to the stochastic interpretation of Timed Automata (STA) \cite{david2015uppaal} and they are as follows, where $N$ and $bound$ indicate the number of simulations to be performed  and the time bound on the simulations respectively:
\begin{enumerate}
\item \emph{Probability Estimation} estimates the probability of a requirement property $\phi$ being satisfied for a given STA model within the time bound: $Pr[bound]\ \phi$.
\item \emph{Hypothesis Testing} checks if the probability of $\phi$ being satisfied is larger than or equal to a certain probability $P_0$: $Pr[bound]\ \phi \ \geqslant\ P_0$.
\item \emph{Probability Comparison} compares the probabilities of two properties being satisfied in certain time bounds: $Pr[bound_1]$ $\phi_1$ $\geqslant$ $Pr[bound_2]$ $\phi_2$.
\item \emph{Expected Value} evaluates the minimal or maximal value of a clock or an integer value while \smc\ checks the STA model: $E[bound; N](min:\phi)$ or $E[bound; N](max:\phi)$.
\item \emph{Simulations}: \smc\ runs $N$ simulations on the STA model and monitors $k$ (state-based) properties/expressions $\phi_1,..., \phi_k$ along the simulations within simulation bound $bound$: $simulate$ $N$ $[\leqslant$ $bound]\{\phi_1,..., \phi_k\}$.
\end{enumerate}

\chapter{Running Example: Traffic Sign Recognition Vehicle}
\label{sec:case-study}

\begin{figure}[t]
\centerline{{\includegraphics[width=5.5in]{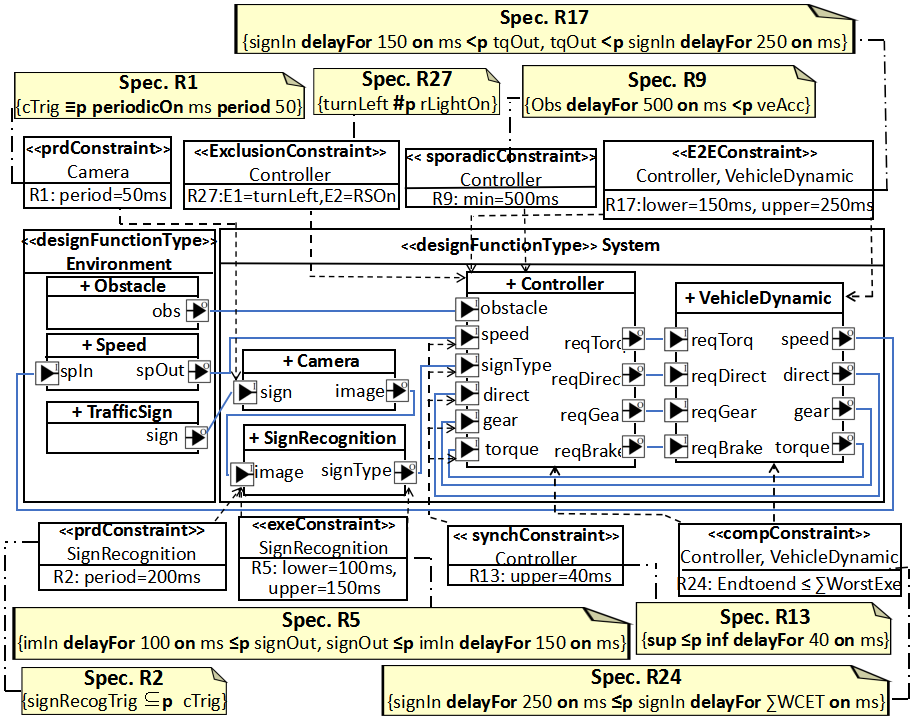}
}}
\caption{AV in \ed\ augmented with \tdl\ constraints ({\gt{R. IDs}}) specified in Pr\ccsl\ ({\gt{Spec. R. IDs}})}
\label{fig:AV_model}
\hfil
\end{figure}

An autonomous vehicle (\av) \cite{mvv,sscps} application using Traffic Sign Recognition is adopted to illustrate our approach. The \av\ reads the road signs, e.g., ``speed limit'' or ``right/left turn'', and adjusts speed and movement accordingly. The functionality of \av, augmented with timing constraints and viewed as {\gt{Functional Design Architecture (FDA)}} ({\gt{designFunctionTypes}}), consists of the following \fp s in Fig. \ref{fig:AV_model}: {\gt{System}} function type contains four \fp s, i.e.,  the {\gt{Camera}} captures sign images and relays the images to {\gt{SignRecognition}} periodically. {\gt{SignRecognition}} analyzes each frame of the detected images and computes the desired images (sign types). {\gt{Controller}} determines how the speed of the vehicle is adjusted based on the sign types and the current speed of the vehicle. {\gt{VehicleDynamic}} specifies the kinematics behaviors of the vehicle. {\gt{Environment}} function type consists of three \fp s, i.e., the information of traffic signs, random obstacles, and speed changes caused by environmental influence described in {\gt{TrafficSign}}, {\gt{Obstacle}}, and  {\gt{Speed}} \fp s respectively.

We consider the {\gt{Periodic}}, {\gt{Execution}}, {\gt{End-to-End}}, {\gt{Synchronization}}, {\gt{Sporadic}}, and {\gt{Comparison}} timing constraints on top of the \av\ \ed\ model, which are sufficient to capture the constraints described in Fig. \ref{fig:AV_model}. Furthermore, we extend \ed/\tdl\ with an {\gt{Exclusion}} timing constraint (R27 -- R31) that integrates relevant concepts from the \ccsl\ constraint, i.e., two events cannot occur simultaneously.

\vspace{0.1in}
\noindent R1. The camera must capture an image every 50ms. In other words, a {\gt{Periodic}}
acquisition of {\gt{Camera}} must be carried out every 50ms.

\noindent R2. The captured image must be recognized by an AV every 200ms, which can be interpreted as a {\gt{Periodic}} constraint on {\gt{SignRecognition}} \fp.

\noindent  R3. The obstacle will be detected by vehicle every 40ms, i.e., a {\gt{Periodic}} timing constraint should be applied on the \emph{obstacle} input port of {\gt{Controller}}.

\noindent R4. The speed of the vehicle should be updated periodically with the period as 30ms, i.e., a {\gt{Periodic}} timing constraint should be applied on the \emph{speed} input port of {\gt{Controller}}.

\noindent R5. The detected image should be computed within [100, 150]ms in order to generate the desired sign type, the {\gt{SignRecognition}} must complete its execution within [100, 150]ms.

\noindent R6. After the  {\gt{Camera}} is triggered, the captured image should be sent out from {\gt{Camera}} within 20 -- 30ms, i.e., the execution time of {\gt{Camera}} should be between 20 and 30ms.

\noindent R7. After an obstacle is detected, the {\gt{Controller}} should send out a request to brake the vehicle within 100 -- 150ms, i.e., the execution time for {\gt{Controller}} should be in the range [100, 150]ms.

\noindent R8. After the command/request from controller is arrived at {\gt{VehicleDynamic}}, the speed should be updated within 50 -- 100ms. That is, the {\gt{Execution}} timing constraint applied on {\gt{VehicleDynamic}} is 50 -- 100ms.

\noindent R9. If the mode of AV switches to ``emergency stop'' due to the certain obstacle, it should not revert back to ``automatic running'' mode within a specific time period. That is interpreted as a {\gt{Sporadic}} constraint, i.e., the mode of AV is changed to ``stop'' because of the encounter of obstacle, it should not revert back to ``run'' mode within 500ms.

\noindent R10. If the mode of AV switches to ``emergency stop'' due to the certain obstacle, it should not revert back to ``accelerate '' mode within a specific time period. That is interpreted as a {\gt{Sporadic}} constraint, i.e., the mode of AV is changed to ``stop'' because of the encounter of obstacle, it should not revert back to ``accelerate'' mode within 500ms.

\noindent R11. If the mode of AV switches to ``emergency stop'' due to the certain obstacle, it should not revert back to ``turn left'' mode within a specific time period. That is interpreted as a {\gt{Sporadic}} constraint, i.e., the mode of AV is changed to ``stop'' because of the encounter of obstacle, it should not revert back to ``turn left'' mode within 500ms.

\noindent  R12. If the mode of AV switches to ``emergency stop'' due to the certain obstacle, it should not revert back to ``turn right'' mode within a specific time period. That is interpreted as a {\gt{Sporadic}} constraint, i.e., the mode of AV is changed to ``stop'' because of the encounter of obstacle, it should not revert back to ``turn right'' mode within 500ms.

\noindent  R13. The required environmental information should arrive to the controller within 40ms. That is input signals ({\gt{speed}}, {\gt{signType}}, {\gt{direct}}, {\gt{gear}} and {\gt{torque}} ports) must be detected by {\gt{Controller}} within a given time window, i.e., the tolerated maximum constraint is 40ms.

\noindent  R14. After the execution of {\gt{Controller}} is finished, all the requests of controller should be updated within 30ms. That is output signals (on {\gt{reqTorq}}, {\gt{reqDirect}}, {\gt{reqGear}}, {\gt{reqBrake}} ports) must be sent within a given time window, i.e., the tolerated maximum constraint is 30ms.

\noindent  R15. The requests from the controller should be arrived to {\gt{VehicleDynamic}}  within 30ms. That is input signals ({\gt{reqTorq}}, {\gt{reqDirect}}, {\gt{reqGear}}, {\gt{reqBrake}}) must be detected by {\gt{VehicleDynamic}} within a given time window, i.e., the tolerated maximum constraint is 30ms.

\noindent  R16. After execution of {\gt{VehicleDynamic}} is finished, the information of vehicle should be updated within 40ms, i.e., the {\gt{Synchronization}} applied on the output ports ({\gt{speed}}, {\gt{direct}}, {\gt{gear}}, {\gt{torque}}) is 40ms.

\noindent  R17. When a traffic sign is recognized, the speed of AV should be updated within [150, 250]ms. An {\gt{End-to-End}} constraint on {\gt{Controller}} and {\gt{VehicleDynamic}}, i.e., the time interval measured from the input arrival of {\gt{Controller}} to the instant at which the corresponding output is sent out from {\gt{VehicleDynamic}} must be within [150, 250]ms.

\noindent  R18. After the camera is triggered to capture the image, the computation of the traffic sign should be finished within [120, 180]ms, i.e., the {\gt{End-to-End}} timing constraint applied on  {\gt{Camera}} and  {\gt{SignRecognition}} should be between 120ms and 180ms.

\noindent  R19. The time interval measured from the instant at which the camera captures an image of traffic sign, to the instant at which the status of AV (i.e., speed, direction) is updated, should be within [270, 430]ms. That is, {\gt{End-to-End}} timing constraint applied on {\gt{Camera}} and {\gt{VehicleDynamic}} should be between 270 and 430ms.

\noindent  R20. When a left turn sign is recognized, the vehicle should turn towards left within 500ms, which can be interpreted as an {\gt{End-to-End}} timing constraint applied on the event {\gt{DetectLeftSign}} and {\gt{StartTurnLeft}}.

\noindent  R21. When a right turn sign is recognized, the vehicle should turn towards right within 500ms, which can be interpreted as an {\gt{End-to-End}} timing constraint applied on the event {\gt{DetectRightSign}} and {\gt{StartTurnRight}}.

\noindent  R22. When a stop sign is recognized, the vehicle should start to brake within 200ms, which can be interpreted as an {\gt{End-to-End}} timing constraint applied on the event {\gt{DetectStopSign}} and {\gt{StartBrake}}.

\noindent  R23. When a stop sign is recognized, the vehicle should be stop completely within 3000ms, which can be interpreted as an {\gt{End-to-End}} timing constraint applied on the event {\gt{DetectStopSign}} and {\gt{Stop}}.

\noindent  R24. The execution time interval from {\gt{Controller}} to {\gt{VehicleDynamic}} should be less than or equal to the sum of the worst case execution time interval of each \fp.

\noindent  R25. The execution time interval from {\gt{Camera}} to {\gt{SignRecognition}} should be less than or equal to the sum of the worst case execution time interval of each \fp.

\noindent  R26. The execution time interval from {\gt{Camera}} to {\gt{VehicleDynamic}}  should be less than or equal to the sum of the worst case execution time interval of each \fp.

\noindent  R27. While AV turns left, the ``turning right'' mode should not be activated. The events of turning left and right considered as exclusive and specified as an {\gt{Exclusion}} constraint.

\noindent  R28. While AV is braking, the ``accelerate'' mode should not be activated. The events of braking and accelerating are considered as exclusive and specified as an {\gt{Exclusion}} constraint.

\noindent  R29. When AV is in the emergency mode because of the obstacle occurrence, ``turn left'' mode must not be activated, i.e., the events of handling emergency and turning left are exclusive and specified as a {\gt{Exclusion}} constraint.

\noindent  R30. When AV is in the emergency mode because of the encounter of an obstacle, ``turn right'' mode must not be activated, i.e., the events of handling emergency and turning right are exclusive and specified as an {\gt{Exclusion}} constraint.

\noindent  R31. When AV is in the emergency mode because of the encounter of an obstacle,  ``accelerate'' mode must not be activated, i.e., the events of handling emergency and accelerating are exclusive and specified as an {\gt{Exclusion}} constraint.

{\gt{Delay}} constraint gives duration bounds (minimum and maximum)
between two events \emph{source} and \emph{target}.  This is specified using \emph{lower, upper}
values given as either {\gt{Execution}} constraint (R5 -- R8) or {\gt{End-to-End}} constraint (R17 -- R23). {\gt{Synchronization}} constraint (R13 -- R16) describes how tightly the
occurrences of a group of events follow each other. All events must
occur within a sliding window, specified by the \emph{tolerance}
attribute, i.e., the maximum time interval allowed between events. {\gt{Periodic}} constraint  states that the period of successive
occurrences of a single event must have a time interval (R1 -- R4). {\gt{Sporadic}} constraint states that \emph{events} can arrive at arbitrary points in time, but with
defined minimum inter-arrival times between two consecutive occurrences (R9 -- R12). {\gt{Comparison}} constraint delimits that two consecutive occurrences of an event should have a minimum inter-arrival time (R24 -- R26). {\gt{Exclusion}} constraint refers that two events must not occur at the same time (R27 -- R31). Those timing constraints are formally specified (see as {\gt{R. IDs}} in Fig.  \ref{fig:AV_model}) using the subset of clock \emph{relations} and \emph{expressions} (see Chapter \ref{sec:preliminary}) in the context of WH. The timing constraints are then verified utilizing \smc\ and are described further in the following chapters.

\chapter{Probabilistic Extension of \emph{Relation} in CCSL}
\label{sec:def_ccsl}
To perform the formal specification and probabilistic verification of \ed\ timing constraints (R1 -- R31 in Sec 3.), \ccsl\ \emph{relations} are augmented with probabilistic properties, called Pr\ccsl,  based on WH \cite{Bernat2001Weakly}. More specifically, in order to describe the bound on the number of permitted timing constraint violations in WH, we extend \ccsl\ \emph{relations} with a probabilistic parameter $p$, where  $p$ is the probability threshold. Pr\ccsl\ is satisfied if and only if the probability of \emph{relation} constraint being satisfied is greater than or equal to $p$.
As illustrated in Fig.  \ref{fig:AV_model}, \ed/\tdl\ timing constraints (R. IDs in Fig.  \ref{fig:AV_model}) can be specified (Spec. R. IDs) using the Pr\ccsl\ \emph{relations} and the conventional \ccsl\ \emph{expressions}.

A \emph{time system} is specified by a set of clocks and clock constraints.  An execution of the time system is a {\gt{run}} where the occurrences of events are clock ticks.

\vspace{-0.05in}
\begin{mydef}[{\gt{Run}}]
A {\gt{run}} $R$ consists of a finite set of consecutive steps where a set of clocks tick at each step $i$. The set of clocks ticking at step $i$ is denoted as $R(i)$, i.e., for all $i$, 0 $\leqslant$ $i$  $\leqslant$ $n$, $R(i) \in R$, where $n$ is the number of steps of $R$.
\end{mydef}

\begin{figure}[htbp]
\centerline{{\includegraphics[width=3.62in]{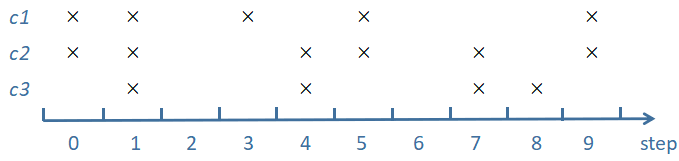}
}}
\caption{Example of {a \gt{Run}}}
\label{fig:example}
\hfil
\end{figure}

\noindent Fig. \ref{fig:example} presents a {\gt{run}} $R$ consisting of $10$ steps and three clocks $c1$, $c2$ and $c3$. The ticks of the three clocks along with steps are shown as ``cross'' symbols (x). For instance, $c1$, $c2$ and $c3$ tick at the first step, hence $R(1)$ = \{$c1,\ c2,\ c3 $\}.

The history of a clock $c$ presents the number of times the clock $c$ has ticked prior to the current step.

\begin{mydef}[{\gt{History}}]
For $c$ $\in$ $C$,  the {\gt{history}} of $c$ in a run $R$ is a function: $H_{R}^c$: $\mathbb{N} \rightarrow \mathbb{N}$. For all instances of step $i$, $i \in \mathbb{N}$,  $H_{R}^c(i)$ indicates the number of times the clock $c$ has ticked prior to step $i$ in run R, which is initialized as 0 at step 0. It is defined as:

\begin{equation*}
H_{R}^c(i) =
\left\{
             \begin{array}{lr}
             0, & i = 0 \\
             H_{R}^c(i-1), & c \notin {R}(i) \wedge i > 0\\
             H_{R}^c(i-1)+1, & c \in {R}(i) \wedge i > 0
             \end{array}
\right.
\end{equation*}
\end{mydef}

\begin{mydef}[{\gt{PrCCSL}}]
Let $c1$, $c2$ and $R$ be two logical clocks and a {\gt{run}}.
The probabilistic extension of relation constraints, denoted  $c1\textcolor{red}{\sim_{p}}c2$, is satisfied if the following condition holds:
\[R \vDash c1\textcolor{red}{\sim_{p}}c2 \Longleftrightarrow {Pr}(c1 \textcolor{red}{\sim} c2) \geqslant p\]
where \textcolor{red}{$\sim$} $\in \{\textcolor{red}{\subseteq, \equiv, \prec, \preceq, \#} \}$,  ${Pr}(c1 \textcolor{red}{\sim} c2)$ is the probability of the relation $c1 \textcolor{red}{\sim} c2$ being satisfied, and  $p$ is the probability threshold.
\end{mydef}

The five \ccsl\ \emph{relations}, {\gt{subclock}}, {\gt{coincidence}}, {\gt{exclusion}},  {\gt{causality}} and {\gt{precedence}}, are considered and their probabilistic extensions are defined.

\vspace{-0.05in}
\begin{mydef}[{\gt{Probabilistic Subclock}}] Let $c1$, $c2$ and $\mathcal{M}$  be two logical clocks and a system model. Given $k$ {\gt{runs}} $=$ $\{R_1, \ldots, R_k \}$,
the probabilistic extension of {\gt{subclock}} relation between $c1$ and $c2$, denoted $c1 \textcolor{red}{\subseteq_p} c2$, is satisfied if the following condition holds:
\[\mathcal{M} \vDash c1 \textcolor{red}{\subseteq_p} c2 \Longleftrightarrow Pr[c1 \textcolor{red}{\subseteq} c2] \geqslant p\]
where $Pr[c1\textcolor{red}{\subseteq} c2]$  $=$ $\frac{1}{k}$ $\sum\limits_{j=1}^{k}$ $\{R_j \models c1 \textcolor{red}{\subseteq} c2\}$, $R_j$ $\in$  $\{R_1, \ldots, R_k \}$, i.e., the ratio of {\gt{runs}} that satisfies the  {\gt{subclock}} relation out of k {\gt{runs}}.
\end{mydef}

\noindent A {\gt{run}} $R_j$ satisfies the {\gt{subclock}} relation between $c1$ and $c2$
``if $c1$ ticks, $c2$ must tick'' holds at every step $i$ in $R_j$, s.t.,
$(R_j \models c1 \textcolor{red}{\subseteq} c2) \Longleftrightarrow
(\forall i$ $0 \leqslant i  \leqslant n,\ c1 \in {R}(i)\implies c2 \in {R}(i))$.
``$R_j \models c1 \textcolor{red}{\subseteq} c2$'' returns 1 if $R_j$ satisfies $c1 \textcolor{red}{\subseteq} c2$, otherwise it returns 0.

{\gt{Coincidence}} relation delimits that two clocks must always tick at the same step, i.e, if $c1$ and $c2$ are coincident, then $c1$ and $c2$ are subclocks of each other.

\vspace{-0.05in}
\begin{mydef}[{\gt{Probabilistic Coincidence}}] The probabilistic {\gt{coincidence}} relation between $c1$ and $c2$, denoted $c1 \textcolor{red}{\equiv_p} c2$, is satisfied over $\mathcal{M}$ if the following condition holds:
\[\mathcal{M} \vDash c1 \textcolor{red}{\equiv_p} c2 \Longleftrightarrow Pr[c1 \textcolor{red}{\equiv} c2] \geqslant p\]
where $Pr[c1 \textcolor{red}{\equiv} c2]$ $=$  $\frac{1}{k}$ $\sum\limits_{j=1}^{k}$ $\left\{R_j \models c1 \textcolor{red}{\equiv} c2 \right\}$ is determined by the number of {\gt{runs}} satisfying the {\gt{coincidence}} relation out of $k$ {\gt{runs}}.
\end{mydef}

\noindent  A {\gt{run}}, $R_j$ satisfies the {\gt{coincidence}} relation on $c1$ and $c2$ if the assertion holds: $\forall i$, $0 \leqslant i  \leqslant n$, $(c1 \in {R}(i)\implies c2 \in {R}(i)) \wedge \ (c2 \in {R}(i)\implies c1 \in {R}(i))$. In other words, the satisfaction of {\gt{coincidence}} relation is established when the two conditions ``if $c1$ ticks, $c2$ must tick'' and ``if $c2$ ticks, $c1$ must tick'' hold at every step.

The inverse of {\gt{coincidence}} relation is {\gt{exclusion}}, which specifies two clocks cannot tick at the same step.

\vspace{-0.05in}
\begin{mydef}[{\gt{Probabilistic Exclusion}}] For all $k$ {\gt{runs}} over $\mathcal{M}$, the probabilistic {\gt{exclusion}} relation between $c1$ and $c2$, denoted $c1 \textcolor{red}{\#_p} c2$, is satisfied
if the following condition holds:
\[\mathcal{M} \vDash c1 \textcolor{red}{\#_p} c2 \Longleftrightarrow Pr[c1 \textcolor{red}{\#} 2] \geqslant p\]
where $Pr[c1 \textcolor{red}{\#} c2]$ $=$ $\frac{1}{k}$ $\sum\limits_{j=1}^{k} \left\{R_j \models c1 \textcolor{red}{\#} c2 \right\}$ is the ratio of the {\gt{runs}} satisfying the {\gt{exclusion}} relation out of $k$ {\gt{runs}}.
\end{mydef}

\noindent A {\gt{run}}, $R_j$, satisfies the {\gt{exclusion}} relation on $c1$ and $c2$  if $\forall i$, $0 \leqslant i  \leqslant n$, $(c1 \in {R}(i)\implies c2 \notin {R}(i)) \wedge \ (c2 \in {R}(i)\implies c1 \notin {R}(i))$, i.e., for every step, if $c1$ ticks, $c2$ must not tick and vice versa.

The probabilistic extension of {\gt{causality}} and {\gt{precedence}} relations are defined based on the history of clocks.

\vspace{0.05in}
\begin{mydef}[{\gt{Probabilistic Causality}}]
The probabilistic  {\gt{causality}} relation between $c1$ and $c2$ ($c1$ is the cause and $c2$ is the effect),
denoted $c1 \textcolor{red}{\preceq_p} c2$, is satisfied if the following condition holds:
\[\mathcal{M} \vDash c1 \textcolor{red}{\preceq_p} c2 \Longleftrightarrow Pr[c1 \textcolor{red}{\preceq} c2] \geqslant p\]
where $Pr[c1 \textcolor{red}{\preceq} c2]$  $=$ $\frac{1}{k}$ $\sum\limits_{j=1}^{k} \left\{R_j \models c1 \textcolor{red}{\preceq} c2 \right\}$, i.e., the ratio of {\gt{runs}} satisfying the {\gt{causality}} relation among the total number of $k$ {\gt{runs}}.
\end{mydef}

\noindent A {\gt{run}} $R_j$ satisfies the {\gt{causality}} relation on $c1$ and $c2$ if
the condition holds: $\forall i$, $0 \leqslant i  \leqslant n$, $H^{c1}_{R}(i) \geqslant H^{c2}_{R}(i)$. A tick of $c1$ satisfies {\gt{causality}} relation if $c2$ does not occur prior to $c1$, i.e., the history of $c2$ is less than or equal to the history of $c1$ at the current step $i$.

The strict {\gt{causality}}, called {\gt{precedence}}, constrains that one clock must always tick faster than the other.

\vspace{0.05in}
\begin{mydef}[{\gt{Probabilistic Precedence}}]
The probabilistic {\gt{precedence}} relation between $c1$ and $c2$, denoted $c1 \textcolor{red}{\prec_p} c2$, is satisfied if the following condition holds:
\[\mathcal{M} \vDash c1 \textcolor{red}{\prec_p} c2 \Longleftrightarrow Pr[c1 \textcolor{red}{\prec} c2] \geqslant p\]
where $Pr[c1 \textcolor{red}{\prec} c2]$  $=$ $\frac{1}{k}$ $\sum\limits_{j=1}^{k} \left\{R_j \models c1 \textcolor{red}{\prec} c2 \right\}$ is determined by the number of  {\gt{runs}} satisfying the {\gt{precedence}} relation out of the $k$ {\gt{runs}}.
\end{mydef}

\noindent  A {\gt{run}} $R_j$ satisfies the {\gt{precedence}} relation if the condition (expressed as $(1) \land (2)$) holds: $\forall i$, $0 \leqslant i  \leqslant n$,
\[
\underbrace{(H^{c1}_{R}(i) \geqslant H^{c2}_{R}(i))}_\text{(1)} \wedge \underbrace{(H^{c2}_{R}(i)=H^{c1}_{R}(i)) \implies (c2 \notin\ {R}(i))}_\text{(2)}
\]

\noindent (1) The history of $c1$ is greater than or equal to the history of $c2$; (2) $c1$ and $c2$ must not be coincident, i.e., when the history of $c1$ and $c2$ are equal, $c2$ must not tick.

\chapter{Specification of Timing Constraints in PrCCSL}
\label{sec:specification}
To describe the property that a timing constraint is satisfied with the probability greater than or equal to a given threshold,  \ccsl\ and its extension Pr\ccsl\ are employed to capture the semantics of probabilistic timing constraints in the context of WH.
Below, we show the \ccsl/Pr\ccsl\ specification of \ed\ timing constraints, including {\gt{Execution}}, {\gt{Periodic}}, {\gt{End-to-End}}, {\gt{Sporadic}}, {\gt{Synchroniza-\\tion}}, {\gt{Exclusion}} and {\gt{Comparison}} timing constraints.
In the system, events are represented as clocks with identical names. The ticks of clocks correspond to the occurrences of the events.

\vspace{0.1in}
\noindent \textbf{Periodic} timing constraints (R1 -- R4) can be specified using {\gt{periodicOn}} \emph{expression} and {\gt{probabilistic coincident}} \emph{relation}. R1 states that the camera must be triggered periodically with a period 50ms. We first construct a periodic clock $prd\_50$ which ticks after every $50$ ticks of $ms$ (the universal clock). Then the property that the {\gt{periodic}} timing constraint is satisfied with probability no less than the threshold $p$ can be interpreted as the \gt{probabilistic coincidence} \emph{relation} between $cmrTrig$ (the event that {\gt{Camera}} \fp\ being triggered) and $prd\_50$. The corresponding specification is given below, where $\triangleq$ means ``is defined as'':
	\begin{equation}
    prd\_50\ \triangleq\ \gt{periodicOn}\ ms\ \gt{period}\ 50
	\end{equation}
	\begin{equation}
	{cmrTrig}\ \textcolor{red}{\equiv_p}\ prd\_50
	\end{equation}
By combining (1) and (2), we can obtain the the specification of R1:
	\begin{equation}
	{cmrTrig}\ \textcolor{red}{\equiv_p}\ \{\gt{periodicOn}\ ms\ \gt{period}\ 50\}
	\end{equation}
In similar, the \ccsl/Pr\ccsl\ specification of R2 -- R4 can be derived:
	\begin{equation}
	\textbf{R2}:\ {signTrig}\ \textcolor{red}{\equiv_p}\ \{\gt{periodicOn}\ ms\ \gt{period}\ 200\}
	\end{equation}
	\begin{equation}
	\textbf{R3}:\ {obsDetect}\ \textcolor{red}{\equiv_p}\ \{\gt{periodicOn}\ ms\ \gt{period}\ 40\}
	\end{equation}
	\begin{equation}
	\textbf{R4}:\ {spUpdate}\ \textcolor{red}{\equiv_p}\ \{\gt{periodicOn}\ ms\ \gt{period}\ 30\}
	\end{equation}
where $signTrig$ is the event/clock that {\gt{SignRecognition}} \fp\ is triggered, $obsDetect$ represents the event that the object detection is activated by the vehicle and $spUpdate$ denotes the event that the speed is updated (i.e., recieved by  {\gt{Controller}}) from the environment.

\noindent Since the $period$ attribute of the \gt{Periodic} timing constraint R2 is 200ms, which is an integral multiple of the $period$ of R1,  R2 can be interpreted as a {\gt{subclock}} relation, i.e., the event $signTrig$ should be a \emph{subclock} of $cmrTrig$. The specification is given below:
	\begin{equation}
	\ {signTrig}\ \textcolor{red}{\subseteq_p}\ cmrTrig
	\end{equation}

\vspace{0.1in}
\noindent \textbf{{Execution}} timing constraints (R5 -- R8) can be specified using {\gt{delayFor}} \emph{expression} and {\gt{probabilistic causality}} \emph{relation}. To specify R5, which states that the {\gt{SignRecognition}} \fp\ must finish execution within [100, 150]ms, i.e., the interval measured from the input event of the \fp\ (i.e., the event that the image is received by the \fp, denoted $imIn$) to the output event of the \fp\ (denoted $imIn$) must have a minimum value 100 and a maximum value 150. We divide this property into two subproperties:
R5(1) The time duration between $imIn$ and  $signOut$ should be greater than 100ms.
R5(2) The time duration between $imIn$ and  $signOut$ should be less than 150ms.
To specify property R5(1), we first construct a new clock $imIn\_dly100$ by delaying $imIn$ (the input event of {\gt{SignRecognition}}) for 100ms. To check whether R5(1) is satisfied within a probability threshold is to verify whether the {\gt{probabilistic causality}} between $imIn\_dly100$ and $signOut$ is valid. The specification of R5(1) is given below:
	\begin{equation}
    imIn\_dly100\ \triangleq\ imIn\ \gt{delayFor}\  100\ \gt{on}\ ms
	\end{equation}
	\begin{equation}
	{imIn\_dly100}\ \textcolor{red}{\preceq_p}\ signOut
	\end{equation}
By combining (7) and (8), we can obtain the the specification of R5(1):
	\begin{equation}
	\{ {imIn}\ \gt{delayFor}\ 100\ \gt{on}\ ms\}\ \textcolor{red}{\preceq_p}\ signOut
	\end{equation}

Similarly, to specify property R5(2), a new clock $imIn\_dly150$ is generated by delaying $imIn$ for 150 ticks on $ms$. Afterwards, the property that R5(2) is satisfied with a probability greater than or equal to $p$ relies on whether the {\gt{probabilistic causality}} \emph{relation} is satisfied. The specification is illustrated as follows:
	\begin{equation}
    imIn\_dly150\ \triangleq\ imIn\ \gt{delayFor}\  150\ \gt{on}\ ms
	\end{equation}
	\begin{equation}
	{signOut}\ \textcolor{red}{\preceq_p}\ mIn\_dly150
	\end{equation}
By combining (10) and (11), we can obtain the the specification of R5(2):
	\begin{equation}
	{signOut}\ \textcolor{red}{\preceq_p}\ \{ {imIn}\ \gt{delayFor}\ 150\ \gt{on}\ ms\}
	\end{equation}
    Analogously, the \ccsl/Pr\ccsl\ specification of R6 -- R8 can be derived:\\
	\begin{equation}
    \begin{split}
     \textbf{R6}:\ \{ {cmrTrig}\ \gt{delayFor}\ 20\ \gt{on}\ ms\}\ \textcolor{red}{\preceq_p}\ cmrOut\\
     {cmrOut}\ \textcolor{red}{\preceq_p}\ \{ {cmrTrig}\ \gt{delayFor}\ 30\ \gt{on}\ ms\}
     \end{split}
	\end{equation}
	\begin{equation}
    \begin{split}
    \textbf{R7}:\ \{ {ctrlIn}\ \gt{delayFor}\ 100\ \gt{on}\ ms\}\ \textcolor{red}{\preceq_p}\ ctrlOut\\
     {ctrlOut}\ \textcolor{red}{\preceq_p}\ \{ {ctrlIn}\ \gt{delayFor}\ 150\ \gt{on}\ ms\}
     \end{split}
	\end{equation}
	\begin{equation}
    \begin{split}
    \textbf{R8}:\ \{ {vdIn}\ \gt{delayFor}\ 50\ \gt{on}\ ms\}\ \textcolor{red}{\preceq_p}\ vdOut\\
     {vdOut}\ \textcolor{red}{\preceq_p}\ \{ {vdIn}\ \gt{delayFor}\ 100\ \gt{on}\ ms\}
     \end{split}
	\end{equation}
where $cmrTrig$ is the event that the {\gt{Camera}} \fp\ being triggered, $cmrOut$ represents the event that the captured image is sent out. $ctrlIn$ ($ctrlOut$) represents the input (resp. output) event of  {\gt{Controller}} \fp. $vdIn$ ($vdOut$) represents the input (resp. output) event of  {\gt{VehicleDynamic}} \fp.

\vspace{0.1in}
\noindent \textbf{Sporadic} timing constraints (R9 -- R12) can be specified using {\gt{delayFor}} \emph{expression} and {\gt{probabilistic precedence}} \emph{relation}. R9 states that there should be a minimum delay between the event $veRun$ (the event that the vehicle is in the ``run'' mode) and the event $obstc$ (the event that the vehicle detects an obstacle), which is specified as 500ms.
To specify R9, we first build a new clock $obstc\_dly500$ by delaying $obstc$ for 500 ticks of $ms$. We then check the {\gt{probabilistic precedence}} relation between $obst\_dly500$ and $veRun$:
	\begin{equation}
    obstc\_dly500\ \triangleq\ obstc\ \gt{delayFor}\  500\ \gt{on}\ ms
	\end{equation}
	\begin{equation}
	{ obstc\_dly500}\ \textcolor{red}{\prec_p}\ veRun
	\end{equation}
By combining (16) and (17), we can obtain the the specification of R9:
	\begin{equation}
	\{obstc\ \gt{delayFor}\  500\ \gt{on}\ ms\}\ \textcolor{red}{\prec_p}\ veRun
	\end{equation}
    Analogously, the \ccsl/Pr\ccsl\ specification of R10 -- R12 can be derived:\\
	\begin{equation}
	\textbf{R10}:\ \{obstc\ \gt{delayFor}\  500\ \gt{on}\ ms\}\ \textcolor{red}{\prec_p}\ veAcc
	\end{equation}
	\begin{equation}
	\textbf{R11}:\ \{obstc\ \gt{delayFor}\  500\ \gt{on}\ ms\}\ \textcolor{red}{\prec_p}\ tLeft
	\end{equation}
	\begin{equation}
	\textbf{R12}:\ \{obstc\ \gt{delayFor}\  500\ \gt{on}\ ms\}\ \textcolor{red}{\prec_p}\ tRight
	\end{equation}
where $veAcc$ is the event/clock that the vehicle is accelerating. $tLeft$ and $tRight$ represent the event that the vehicle transits from the ``emergency stop'' mode to ``turn left'' and ``turn right'' mode respectively.

\vspace{0.1in}
\noindent \textbf{Synchronization} timing constraints (R13 -- R16) can be specified using {\gt{infimum}} and {\gt{supremum}} \emph{expression}, together with {\gt{probabilistic precedence}} \emph{relation}.
R13 states that the five input events must be detected by {\gt{Controller}}
within the maximum tolerated time, given as 40ms.
The synchronization timing constraint can be interpreted as: the time interval between the earliest/fastest and the latest/slowest event among the five input events, i.e.,  {\emph{speed}}, {\emph{signType}}, {\emph{direct}}, {\emph{gear}} and {\emph{torque}},  must not exceed 40ms.
To specify the constraints, {\gt{infimum}} is utilized to express
the fastest event (denoted $inf_{ctrlIn}$) while {\gt{supremum}} is utilized to specify the slowest event $sup_{ctrlIn}$. $sup_{ctrlIn}$ and $inf_{ctrlIn}$ are defined as:
\begin{equation}
sup_{ctrl}\ \triangleq\ {{\gt{Sup}}}({{\gt{Sup}}}({\emph{speed}},\ {\emph{signType}}),\ {{\gt{Sup}}}({ \gt{Sup}}({\emph{direct}},\ {\emph{gear}}),\ {\emph{torque}}))
\end{equation}
\begin{equation}
inf_{ctrl}\ \triangleq\ {\gt{Inf}}({\gt{Inf}}({\emph{speed}},\ {\emph{signType}}),\ {\gt{Inf}}({\gt{Inf}}({\emph{direct}},\ {\emph{gear}}),\ {\emph{torque}}))
\end{equation}
where {\gt{Inf}}($c1$, $c2$) (resp. {\gt{Sup}}($c1$, $c2$)) is the {\gt{infimum}} (resp. {\gt{supremum}}) operator returns the
slowest clock faster than $c1$ and $c2$.
Afterwards, we construct a new clock $inf_{ctrlIn}\_dly40$ that is the $inf_{ctrlIn}$ delayed for 40 ticks of $ms$, which is defined as:
	\begin{equation}
    inf_{ctrlIn}\_dly40\ \triangleq\ inf_{ctrl}\ \gt{delayFor}\  40\ \gt{on}\ ms
	\end{equation}
Therefore, the {\gt{synchronization}} constraint R13 can be represented as the {\gt{proba-\\bilistic causality}} \emph{relation} between $sup_{ctrlIn}$ and $inf_{ctrlIn}\_dly40$, given as the \ccsl/Pr\ccsl\ expression below:
	\begin{equation}
    sup_{ctrlIn}\ \textcolor{red}{\preceq_p}\  {inf_{ctrlIn}\_dly40}
	\end{equation}
By combining (24) and (25), we can obtain the the specification of R13:
	\begin{equation}
    sup_{ctrlIn}\ \textcolor{red}{\preceq_p}\  \{inf_{ctrlIn}\ \gt{delayFor}\  40\ \gt{on}\ ms\}
	\end{equation}

\noindent In similar, the \ccsl/Pr\ccsl\ specification of R14 -- R16 can be derived.
For R14, we first construct the clocks that represent the fastest and slowest output event/clock among the four output events of {\gt{Controller}} \fp, i.e., \emph{reqTorq}, \emph{reqDirect}, \emph{reqGear} and \emph{reqBrake}. Then the property that the synchronization constraint is satisfied with a probability greater than or equal to $p$ can be interpreted as a {\gt{probabilistic causality}} \emph{relation}:
	\begin{equation}
    \begin{split}
	\textbf{R14}:
    sup_{ctrlOut}\ \triangleq\ \gt{Sup}(\gt{Sup}(\emph{reqTorq},\ \emph{reqDirect}),\ \gt{Sup}(\emph{reqGear},\ \emph{reqBrake}))\\
    inf_{ctrlOut}\ \triangleq\ \gt{Inf}(\gt{Inf}(\emph{reqTorq},\ \emph{reqDirect}),\ \gt{Inf}(\emph{reqGear},\ \emph{reqBrake}))\\
    sup_{ctrlOut}\ \textcolor{red}{\preceq_p}\  \{inf_{ctrlOut}\ \gt{delayFor}\  30\ \gt{on}\ ms\}
    \end{split}
    \end{equation}
For R15, we first construct the fastest and slowest input event/clock among the four input events of {\gt{VehicleDynamic}} , i.e., \emph{reqTorq}, \emph{reqDirect}, \emph{reqGear} and \emph{reqBrake}. Then the property that the synchronization constraint is satisfied with a probability greater than or equal to $p$ can be interpreted as a {\gt{probabilistic causality}} \emph{relation}:
	\begin{equation}
    \begin{split}
	\textbf{R15}:
    sup_{vdIn}\ \triangleq\ \gt{Sup}(\gt{Sup}(\emph{reqTorq},\ \emph{reqDirect}),\ \gt{Sup}(\emph{reqGear},\ \emph{reqBrake}))\\
    inf_{vdIn}\ \triangleq\ \gt{Inf}(\gt{Inf}(\emph{reqTorq},\ \emph{reqDirect}),\ \gt{Inf}(\emph{reqGear},\ \emph{reqBrake}))\\
    sup_{vdIn}\ \textcolor{red}{\preceq_p}\  \{inf_{vdIn}\ \gt{delayFor}\  40\ \gt{on}\ ms\}
    \end{split}
    \end{equation}
For R16, we first construct the fastest and slowest output event/clock among the four output events of {\gt{VehicleDynamic}} , i.e., \emph{speed}, \emph{direct}, \emph{torque} and \emph{gear}. Then the property that the synchronization constraint is satisfied with a probability greater than or equal to $p$ can be interpreted as a {\gt{probabilistic causality}} \emph{relation}:
    \begin{equation}
    \begin{split}
	\textbf{R16}:
    sup_{vdOut}\ \triangleq\ \gt{Sup}(\gt{Sup}(speed,\ direct),\ \gt{Sup}(gear,\ torque))\\
    inf_{vdOut}\ \triangleq\ \gt{Inf}(\gt{Inf}(speed,\ direct),\ \gt{Inf}(gear,\ torque))\\
    sup_{vdOut}\ \textcolor{red}{\preceq_p}\  \{inf_{vdOut}\ \gt{delayFor}\  40\ \gt{on}\ ms\}
    \end{split}
    \end{equation}

\vspace{0.1in}
\noindent \textbf{{End-to-End}} timing constraints (R17 -- R23) can be specified using {\gt{delayFor}} \emph{expression} and {\gt{probabilistic precedence}} \emph{relation}.
To specify R17, which limits that the time duration measured from the instant of the occurrence of the event that {\gt{Controller}} \fp\ receive the traffic sign type information  (denoted as $signIn$),   to the occurrence of event that the speed is sent out from the output port of  {\gt{VehicleDynamic}} \fp\ (denoted as $spOut$) should be between 150 and 250ms.
We divide this property into two subproperties:
R17(1). The time duration between $signIn$ and  $spOut$ should be more than 150ms.
R17(2). The time duration between $signIn$ and  $spOut$ should be less than 250ms.
To specify property R17(1), we first construct a new clock $signIn\_dly150$ by delaying $signIn$  for 150ms. To check whether R17(1) is satisfied within a probability threshold $p$ is to verify whether the {\gt{probabilistic precedence}} between $signIn\_dly150$ and $spOut$ is valid. The specification of R17(1) is given below:
	\begin{equation}
    signIn\_dly150\ \triangleq\ signIn\ \gt{delayFor}\  150\ \gt{on}\ ms
	\end{equation}
	\begin{equation}
	{signIn\_dly150}\ \textcolor{red}{\prec_p}\ spOut
	\end{equation}
By combining (30) and (31), we can obtain the the specification of R17(1):
	\begin{equation}
	\{ {signIn}\ \gt{delayFor}\ 150\ \gt{on}\ ms\}\ \textcolor{red}{\prec_p}\ spOut
	\end{equation}

\noindent Similarly, to specify property R17(2), a new clock $signIn\_dly250$ is generated by delaying $signIn$ for 250 ticks on $ms$. Afterwards, the property that R17(2) is satisfied with a probability greater than or equal to $p$ relies on whether the {\gt{probabilistic precedence}} \emph{relation} is satisfied. The specification is illustrated as follows:
	\begin{equation}
    signIn\_dly250\ \triangleq\ signIn\ \gt{delayFor}\  250\ \gt{on}\ ms
	\end{equation}
	\begin{equation}
	{spOut}\ \textcolor{red}{\prec_p}\  signIn\_dly250
	\end{equation}
By combining (33) and (34), we can obtain the the specification of R17(2):
	\begin{equation}
	{spOut}\ \textcolor{red}{\prec_p}\ \{ {signIn}\ \gt{delayFor}\ 250\ \gt{on}\ ms\}
	\end{equation}
In similar, the \ccsl/Pr\ccsl\ specification of R18 -- R23 can be derived:\\
	\begin{equation}
    \begin{split}
     \textbf{R18}:\ \{ {cmrTrig}\ \gt{delayFor}\ 120\ \gt{on}\ ms\}\ \textcolor{red}{\prec_p}\ signOut\\
     {signOut}\ \textcolor{red}{\prec_p}\ \{ {cmrTrig}\ \gt{delayFor}\ 180\ \gt{on}\ ms\}
     \end{split}
	\end{equation}
	\begin{equation}
    \begin{split}
    \textbf{R19}:\ \{ {cmrTrig}\ \gt{delayFor}\ 270\ \gt{on}\ ms\}\ \textcolor{red}{\prec_p}\ spOut\\
     {spOut}\ \textcolor{red}{\prec_p}\ \{ {cmrTrig}\ \gt{delayFor}\ 430\ \gt{on}\ ms\}
     \end{split}
	\end{equation}
	\begin{equation}
    \textbf{R20}:\ \{ {startTurnLeft}\ \textcolor{red}{\prec_p}\  DetectLeftSign\ \gt{delayFor}\ 500\ \gt{on}\ ms\}
	\end{equation}
	\begin{equation}
    \textbf{R21}:\ \{ {startTurnRight}\ \textcolor{red}{\prec_p}\  DetectRightSign\ \gt{delayFor}\ 500\ \gt{on}\ ms\}
	\end{equation}
	\begin{equation}
    \textbf{R22}:\ \{ {startBrake}\ \textcolor{red}{\prec_p}\  DetectStopSign\ \gt{delayFor}\ 500\ \gt{on}\ ms\}
	\end{equation}
	\begin{equation}
    \textbf{R23}:\ \{{Stop}\ \textcolor{red}{\prec_p}\  DetectStopSign\ \gt{delayFor}\ 3000\ \gt{on}\ ms\}
	\end{equation}

\vspace{0.1in}
\noindent \textbf{Comparison} timing constraints (R24 -- R26) can be specified using {\gt{delayFor}} \emph{expression} and {\gt{probabilistic causality}} \emph{relation}. R24 states that the execution time interval from {\gt{Controller}} to {\gt{VehicleDynamic}} should be less than or equal to the sum of the worst case execution time of {\gt{Controller}} and {\gt{VehicleDynamic}}, denoted as \emph{W$_{ctrl}$} and \emph{W$_{vd}$} respectively.
To specify {\gt{comparison}} constraint, we first construct a new clock $signIn\_dly250$ by delaying $signIn$ for 250 ticks of $ms$. Afterwards, we generate another new clock $signIn\_dlysw$ that is the $signIn$ clock delayed for sum of the \emph{worst case execution time} of the two \fp s. The specification is illustrated as follows:
	\begin{equation}
    signIn\_dly250\ \triangleq\ signIn\ \gt{delayFor}\  250\ \gt{on}\ ms
	\end{equation}
	\begin{equation}
    signIn\_dlysw\ \triangleq\ signIn\ \gt{delayFor}\  (W_{ctrl}+W_{vd})\ \gt{on}\ ms
	\end{equation}

\noindent Therefore, the property that the probability of comparison constraint is satisfied should be greater than or equal to the threshold $p$ can be interpreted as a
 {\gt{probabilistic causality}} \emph{relation} between $signIn\_dly250$ and $ signIn\_dlysw$:
 	\begin{equation}
	 signIn\_dly250\ \textcolor{red}{\preceq_p}\ signIn\_dlysw
	\end{equation}
By combining (42), (43) and (44), we can obtain the the specification of R24:
	\begin{equation}
	\{signIn\ \gt{delayFor}\  250\ \gt{on}\ ms\}\ \textcolor{red}{\preceq_p}\ \{signIn\ \gt{delayFor}\  (W_{ctrl}+W_{vd})\ \gt{on}\ ms\}
	\end{equation}
\noindent Analogously, the \ccsl/Pr\ccsl\ specification of R25 and R26 can be derived:\\
    	\begin{equation}
	\textbf{R25}:\ \{cmrTrig\ \gt{delayFor}\  180\ \gt{on}\ ms\}\ \textcolor{red}{\preceq_p}\ \{cmrTrig\ \gt{delayFor}\  (W_{cmr}+W_{sr})\ \gt{on}\ ms\}
	\end{equation}
	\begin{equation}
    \begin{split}
	\textbf{R26}:\ \{cmrTrig\ \gt{delayFor}\  430\ \gt{on}\ ms\}\ \textcolor{red}{\preceq_p}\ \\ \{cmrTrig\ \gt{delayFor}\ (W_{cmr}+ W_{sr}+ W_{ctrl}+W_{vd})\ \gt{on}\ ms\}
    \end{split}
	\end{equation}
where $W_{cmr}$ and $W_{vd}$ represent the worst case execution time of {\gt{Camera}} and {\gt{SignRecognition}} respectively.

\vspace{0.1in}
\noindent \textbf{Exclusion} timing constraints (R27 -- R31) can be specified using {\gt{exclusion}} \emph{relation} directly. R27 states that the two events $turnLeft$ (the event that the vehicle is turning left) and $rightOn$ (the event that the turn right mode is activated) should be exclusive, which can be expressed as:
	\begin{equation}
    {turnLeft}\ \textcolor{red}{\#_p}\ rightOn
	\end{equation}
Analogously, the {\gt{Exclusion}} timing constraints R28 -- R31 can be specified using {\gt{exclusion}} \emph{relation}:
	\begin{equation}
    \textbf{R28}:\ {veAcc}\ \textcolor{red}{\#_p}\ veBrake
	\end{equation}
	\begin{equation}
    \textbf{R29}:\ {emgcy}\ \textcolor{red}{\#_p}\ turnLeft
	\end{equation}
	\begin{equation}
    \textbf{R30}:\ {emgcy}\ \textcolor{red}{\#_p}\ rightOn
	\end{equation}
	\begin{equation}
    \textbf{R31}:\ {emgcy}\ \textcolor{red}{\#_p}\ veAcc
	\end{equation}
where $emgcy$ is the event that the vehicle is in the \emph{emergency} mode, $veBrake$ and $veAcc$ represent the event that the vehicle is braking or accelerating, respectively.

\chapter{Translating CCSL \& PrCCSL into UPPAAL-SMC}
\label{sec:model}

To formally verify the \ed\ timing constraints given in Chapter 3 using \smc, we investigate how those constraints, specified in \ccsl\ \emph{expressions} and Pr\ccsl\ \emph{relations}, can be translated into STA and probabilistic \smc\ queries \cite{david2015uppaal}.
\ccsl\ \emph{expressions} construct new clocks and the \emph{relations} between the new clocks are specified using Pr\ccsl. We first provide strategies that represent \ccsl\ \emph{expressions} as STA. We then present how the \ed\ timing constraints defined in Pr\ccsl\ can be translated into the corresponding STAs and \smc\
 queries based on the strategies.

\section{Mapping CCSL to UPPAAL-SMC}

\begin{figure}[htbp]
\centering
  \subfigure[ms]{
  \includegraphics[width=0.72in]{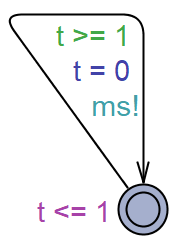}
  \label{fig:TimeUnit}}
  \subfigure[Tick and History]{
  \includegraphics[width=1.75in]{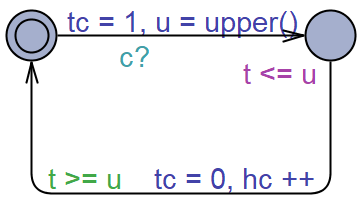}
  \label{fig:tick}}
  \subfigure[Simulation of Tick (tc) and History (hc)]{
  \includegraphics[width=3.0in]{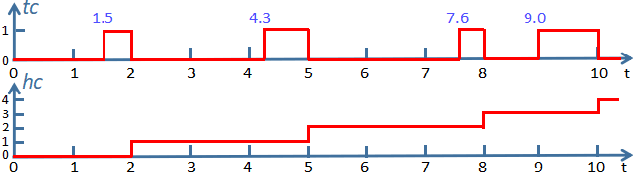}
  \label{fig:tc}}
  \caption{\smc\ model of clock tick and history}
  \label{fig:tickandhis}
\end{figure}

We first describe how the universal clock (TimeUnit $ms$), tick and history of \ccsl\ can be mapped to the corresponding STAs. Using the mapping, we then demonstrate that \ccsl\ \emph{expressions} can be modeled as STAs. The TimeUnit is implicitly represented as a single \emph{step} of time progress in \smc's \emph{clock} \cite{ksac14}. The STA of TimeUnit (universal time defined as $ms$) consists of one location and one outgoing transition whereby the physical time and the duration of TimeUnit $ms$ are represented by the \emph{clock} variable $t$ in Fig. \ref{fig:tickandhis}.(a). \emph{clock} resets every time a transition is taken. The duration of TimeUnit is expressed by the invariant  $t\leqslant 1$, and guard $t\geqslant 1$, i.e., a single step of the discrete time progress (tick) of universal time.

A clock $c$, considered as an event in \smc, and its tick, i.e., an occurrence of the event, is represented by the synchronization channel $c!$. Since \smc\ runs in chronometric semantics, in order to describe the discretized steps of {\gt{run}}s ($R$s), we consider if $c$ ticks in the time range of  $[i, i+1)$ ($i+1$ is excluded), $c$ ticks at step $i$. The STA of tick and history is shown in Fig. \ref{fig:tickandhis}.(b). $hc$ is the history of $c$, and  $tc$ indicates whether $c$ ticks at the current step. A function $upper()$ rounds the time instant (real number) up to the nearest greater integer. When $c$ ticks via $c?$ at the current time step, $tc$ is set to 1 prior to the time of the next step  ($t<u$). $hc$ is then increased by 1 ($hc$++) at the successive step (i.e., when $t=u$). For example, when $c$ ticks at $time = 1.5$  (see Fig. \ref{fig:tickandhis}.(c)), $upper()$ returns the value of 2 and $tc$ becomes 1 during the time interval [1.5, 2), followed by $hc$ being increased by 1 at $t=2$.

Based on the mapping patterns of $ms$, tick and history, we present how {\gt{periodicOn}}, {\gt{delayFor}}, {\gt{infimum}} and {\gt{supremum}} \emph{expressions} can be represented as \smc\ models.

\begin{figure}[htbp]
\centering
  \subfigure[PeriodicOn]{
  \includegraphics[width=1.3in]{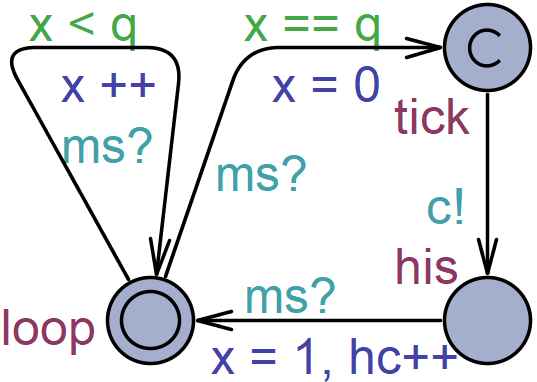}
  \label{fig:periodic}}
  \subfigure[Source]{
  \includegraphics[width=1.1in]{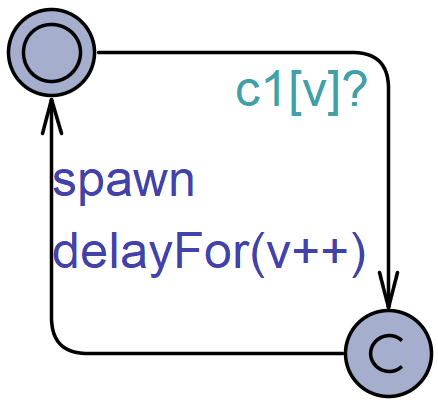}
  \label{fig:source}}
  \subfigure[DelayFor]{
  \includegraphics[width=1.5in]{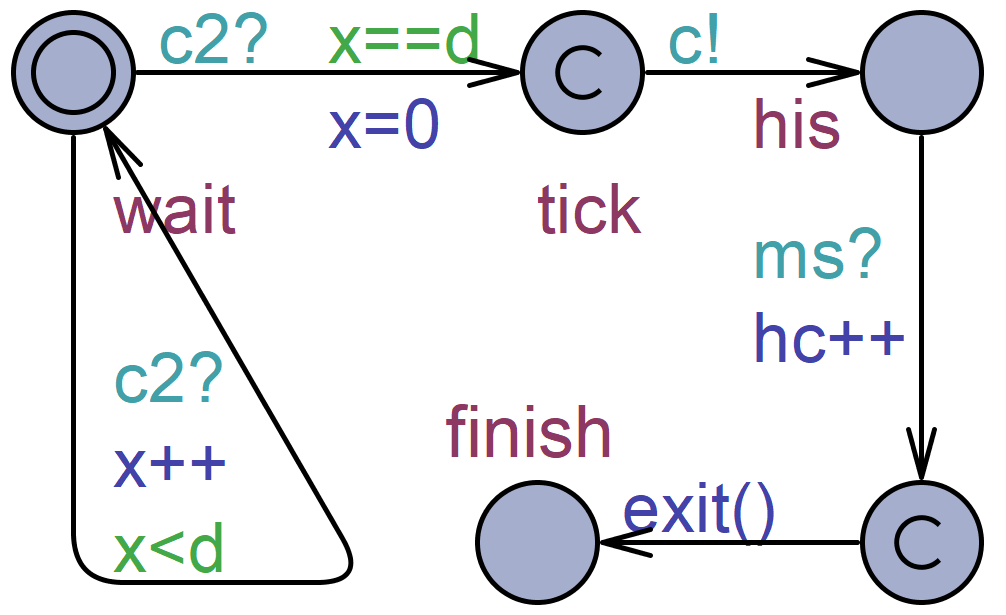}
  \label{fig:delayfor}}
  \subfigure[Infimum]{
  \includegraphics[width=1.6in]{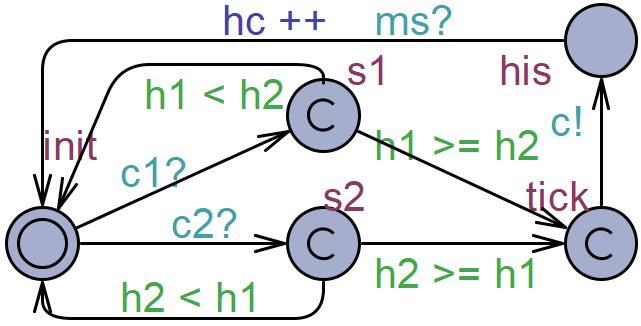}
  \label{fig:inf}}
  \subfigure[Supremum]{
  \includegraphics[width=3.0in]{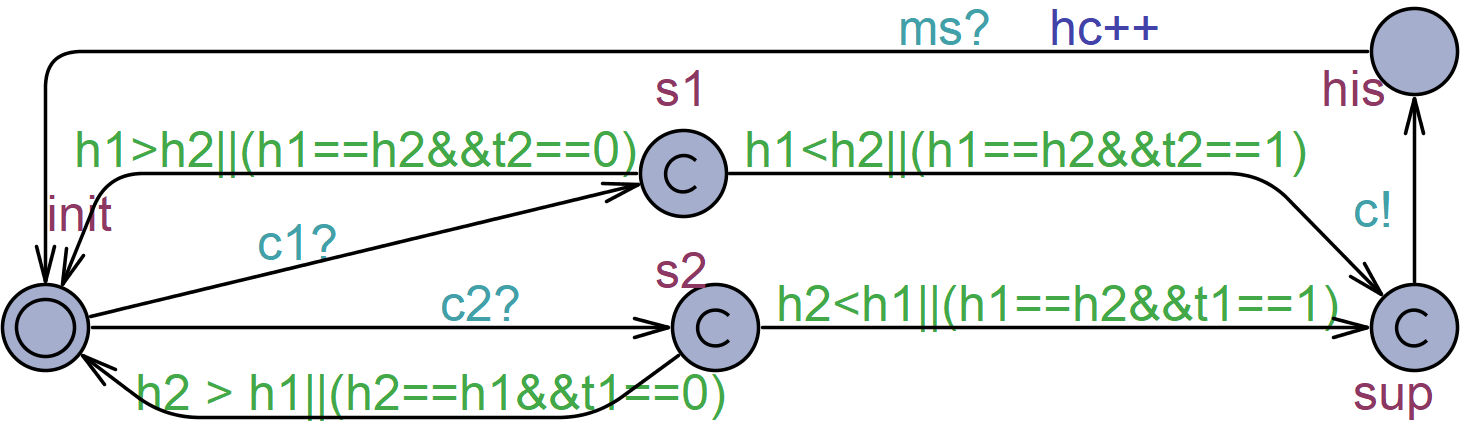}
  \label{fig:sup}}
  \caption{STA of \ccsl\ \emph{expressions}}
\label{fig:CCSL_smc}
\end{figure}

\vspace{0.05in}
\noindent\textbf{PeriodicOn}: $c \ \triangleq\ \ { \gt{periodicOn}}\ ms\ {\gt{period}}\ q$, where $\triangleq$ means ``is defined as''. {\gt{PeriodicOn}} builds a new clock $c$ based on $ms$ and a \emph{period} parameter $q$, i.e., $c$ ticks at every $q^{th}$ tick of $ms$. The STA of {\gt{periodicOn}} is illustrated in Fig. \ref{fig:CCSL_smc}.(a).  This STA initially stays in the \emph{loop} location to detect $q$ occurrences (ticks) of $ms$. The value $x$ counts the number of $ms$ ticks. When $ms$ occurs ($ms?$), the STA takes the outgoing transition and increases $x$ by 1. It ``iterates'' until $ms$ ticks $q$ times ($x==q$), then it activates the tick of $c$ (via $c!$).  At the successive step ($ms?$), it updates the history of $c$ ($hc$++) and sets $x=1$. The STA then returns to \emph{loop} and repeats the calculation. This {\gt{periodicOn}} STA can be used for the translation of \ed\ {\gt{Periodic}} timing constraint (R1 in Fig. \ref{fig:AV_model}) into its \smc\ model.

\vspace{0.05in}
\noindent\textbf{DelayFor}: $c \ \triangleq\ c1\ { \gt{delayFor}}\ d\ {\gt{on}}\ c2$.  {\gt{DelayFor}} defines a new clock $c$ based on $c1$ (\emph{base} clock) and $c2$ (\emph{reference} clock), i.e., each time $c1$ ticks, at the $d^{th}$ tick of $c2$, $c$ ticks (each tick of $c$ corresponds to a tick of $c1$). Kang et al. \cite{ksac14} and Suryadevara et al. \cite{Suryadevara2013Validating} presented translation rules of {\gt{delayFor}} into \uppaal\ models. However, their approaches are not applicable in the case after $c1$ ticks, and $c1$ ticks again before the $d^{th}$ tick of $c2$ occurs. For example (see Fig. \ref{fig:example}), assume that $d$ is 3. After the $1^{st}$ tick of $c1$ (at step 0) happens, if $c1$ ticks again (at step 2) before the $3^{rd}$ tick of $c2$ occurs (at step 4), the $2^{nd}$ tick of $c1$ is discarded in their approaches. To alleviate the restriction, we utilize spawnable STA \cite{david2015uppaal} as semantics denotation of {\gt{delayFor}} expression and the STA of {\gt{delayFor}} is shown in  Fig. \ref{fig:CCSL_smc}.(c).
As presented in Fig. \ref{fig:CCSL_smc}.(b), when the $v^{th}$ tick of $c1$ occurs ($c1[v]?$), its {\gt{delayFor}} STA is spawned by {\gt{source}} STA. The spawned STA stays in the \emph{wait} location until $c2$ ticks $d$ times. When $c2$ ticks $d$ times ($x==d$), it transits to the \emph{tick} location and triggers $c$ ($c!$).
At the next step ($ms?$), the STA increases $hc$ by 1 and moves to $finish$ location and then becomes inactive, i.e., calculation of the $v^{th}$ tick of $c$ is completed. This {\gt{delayFor}} STA can be utilized to construct the \smc\ models of \ed\ timing requirements R5 -- R26 in Chapter 3.

Given two clocks $c1$ and $c2$, their {\gt{infimum}} (resp. {\gt{supremum}}) is informally defined as the slowest (resp. fastest) clock faster (resp. slower) than both $c1$ and $c2$. {\gt{infimum}} and {\gt{supremum}} are useful in order to group events occurring at the same time and decide which one occurs first and which one occurs last. The representative STAs for both \emph{expressions} are utilized for the translation of  \ed\ {\gt{Synchronization}} timing constraint (R13 in Chapter 3) into the \smc\ model.

\vspace{0.05in}
\noindent\textbf{Infimum} creates a new clock $c$, which is the slowest clock faster than $c1$ and $c2$.
The STA of {\gt{infimum}} is illustrated in Fig. \ref{fig:CCSL_smc}.(d). When $c1$ ($c2$) ticks via $c1?$ ($c2?$), the STA transits to the \emph{s1} (\emph{s2}) location and compares the history of the two clocks ($h1$ and $h2$) to check whether the current ticking clock $c1$ ($c2$) is faster than $c2$ ($c1$). If so, i.e., the condition ``$h1$ $\geqslant$ $h2$ ($h2$ $\geqslant$ $h1$)'' holds, the STA takes a transition to the \emph{tick} location and activates the tick of $c$ ($c!$). After updating the history ($hc$++), it returns to the \emph{init} location and repeats the calculation.

\vspace{0.05in}
\noindent\textbf{Supremum} builds a new clock $c$, which is the fastest clock slower than $c1$ and $c2$. It states that if $c1$ ticks at the current step and $c1$ is slower than $c2$, then $c$ ticks.
The STA of {\gt{supremum}} is shown in Fig. \ref{fig:CCSL_smc}.(e). When $c1$ ($c2$) ticks via $c1?$ ($c2?$), the STA transits to the \emph{s1} (\emph{s2}) location and compares the history of the two clocks and decides whether $c1$ ($c2$) is slower than $c2$ ($c1$). If $c1$ ($c2$) ticks slower than $c2$ ($c1$), i.e., $h1<h2$ ($h2<h1$), or $c1$ and $c2$ tick at the same rate, i.e., ``$h1==h2$ $\&\&$ $t_2==1$ ($h1==h2$ $\&\&$ $t_1==1$)'' holds, the tick of $c$ is triggered.  The STA then updates the history of $c$ and goes back to \emph{init} and repeats the process.

\section{Representation of \prccsl\ in UPPAAL-SMC}

In this section, the translation of \ed\ timing constraints specified in Pr\ccsl\ into
STA and \emph{Hypothesis Testing} query (refer to Chapter \ref{sec:preliminary}) is provided from the view point of the analysis engine \smc.

Recall the definition of Pr\ccsl\ in Chapter 4.  The probability of a \emph{relation} being satisfied is interpreted as a ratio of runs that satisfies the \emph{relation} among all runs. It is specified as \emph{Hypothesis Testing} queries in \smc,  $H_0$:  $\frac{m}{k} \geqslant P$ against $H_1$: $\frac{m}{k} < P$, where $m$ is the number of runs satisfying the given \emph{relation} out of all $k$ runs. $k$ is decided by strength parameters $\alpha$ (the probability of false positives, i.e., accepting $H_1$ when $H_0$ holds) and $\beta$  (probability of false negatives, i.e., accepting $H_0$ when $H_1$ holds), respectively \cite{Bulychev2012UPPAAL}.

Based on the mapping patterns of tick and history in Chapter 6.1, the probabilistic extension of {\gt{exclusion}}, {\gt{causality}} and {\gt{precedence}} relations are expressed
as \emph{Hypothesis Testing} queries straightforwardly.

\vspace{0.05in}
\noindent\textbf{Probabilistic Exclusion} is employed to specify \ed\ {\gt{Exclusion}} timing constraint, $turnLeft\ \textcolor{red}{\#_p}\ rightOn$ (Spec. R27 in Fig. \ref{fig:AV_model}).
It states that the two events, \emph{turnLeft} and \emph{rightOn} (the vehicle is turning left and right), must be exclusive. The ticks of $turnLeft$ and $rightOn$ events are modeled using the STA in Fig. \ref{fig:tickandhis}.(b).
Based on the definition of {\gt{probabilistic exclusion}} (Chapter \ref{sec:def_ccsl}), R8 is expressed in \emph{Hypothesis Testing} query: $Pr[bound]$ $([\ ]$$((t_{turnLeft}$ $\implies$ $\neg$ $t_{rightOn})$ $\wedge$ $(t_{rightOn}$ $\implies$ $\neg$ $t_{turnLeft})))$ $\geqslant$ $P$, where $t_{turnLeft}$ and $t_{rightOn}$ indicate the ticks of $turnLeft$ and $rightOn$, respectively. $bound$ is the time bound of simulation, in our setting $bound = 3000$.

\vspace{0.05in}
\noindent\textbf{Probabilistic Causality} is used to specify \ed\ {\gt{Synchronization}} timing constraint, $sup\ \textcolor{red}{\preceq_p}\ \{inf\ {\gt{delayFor}}\ 40\ {\gt{on}}\ ms\}$ (Spec. R13 in Fig. \ref{fig:AV_model}), where \emph{sup} (\emph{inf}) is the fastest (slowest) event slower (faster) than five input events, \emph{speed}, \emph{signType}, \emph{direct}, \emph{gear} and \emph{torque}. Let $\gt{SUP}$ and $\gt{INF}$ denote the {\gt{supremum}} and {\gt{infimum}} operator, i.e., $\gt{SUP}(c1,\ c2)$ (resp. $\gt{INF}(c1,\ c2)$) returns the {\gt{supremum}} (resp. {\gt{infimum}}) of clock $c1$ and $c2$.
\emph{sup} and \emph{inf} can now be expressed with the nested operators (where $\triangleq$ means ``is defined as''):
\[sup\ \triangleq\ \gt{SUP}({{speed}},\ \gt{SUP}(\gt{SUP}({{signType}},\ {{direct}}),\ \gt{SUP}({{gear}},\ {{torque}})))\]
\[inf\ \triangleq\ \gt{INF}({{speed}},\ \gt{INF}(\gt{INF}({{signType}},\ {{direct}}),\ \gt{INF}({{gear}},\ {{torque}})))\]
%

For the translation of $sup$ ($inf$) into \smc\ model,  we employ the STA of {\gt{supremum}} (resp. {\gt{infimum}}) (Fig. \ref{fig:CCSL_smc}.(d) and (e)) for each $\gt{SUP}$ ($\gt{INF}$) operator.
A new clock \emph{dinf} is generated by delaying \emph{inf} for 40 ticks of $ms$: $dinf \triangleq \{inf\ { \gt{delayFor}}\ 40\ {\gt{on}}\ ms\}$. The \smc\ model of \emph{dinf} is achieved by adapting the spawnable \emph{DelayFor} STA (Fig. \ref{fig:CCSL_smc}). Based on the {\gt{probabilistic causality}} definition, R13 is interpreted as: $Pr[\leqslant bound]([\ ]\ h_{sup}\ \geqslant h_{dinf})\ \geqslant\ P$, where $h_{sup}$ and $h_{dinf}$ are the history of \emph{sup} and \emph{dinf} respectively.

Similarly, {\gt{Execution}} (R5) and {\gt{Comparison}} (R25) timing constraints specified in {\gt{probabilistic causality}} using {\gt{delayFor}} can be translated into \emph{Hypothesis Testing} queries. R5 (\{$imIn\ {\gt{delayFor}}\ 100\ {\gt{on}}\ ms\}\ \textcolor{red}{\preceq_p}\ signOut$, $signOut\ \\ \textcolor{red}{\preceq_p}\ \{imIn\ {\gt{delayFor}}\ 150\ {\gt{on}}\ ms\}$) specifies that the execution time of {\gt{SignRecog-\\nition}} \fp\ measured from input port $imIn$ to output port $signOut$ should be limited within [100, 150]ms. To translate {\gt{Execution}} timing constraint into \smc\ STA, two new clocks \emph{SL} and \emph{SU} are constructed by delaying \emph{imIn} for 100 and 150 ticks of \emph{ms}: $SL \triangleq \{imIn\ {\gt{delayFor}}\ 100\ {\gt{on}}\ ms\}$, $SU \triangleq \{imIn\ {\gt{delayFor}}\ 150\ {\gt{on}}\ ms\}$. According to the definition of {\gt{probabilistic causality}}, R5 can be specified as: $Pr[\leqslant bound]([\ ]\ h_{SL}\ \geqslant h_{S})\ \geqslant\ P$, $Pr[\leqslant bound]([\ ]\ h_{S}\ \geqslant h_{SU})\ \geqslant\ P$, where $h_{SU}$ and $h_{SL}$ represent the history of \emph{SU} and \emph{SL}, and $h_{S}$ indicates the history of clock $signOut$.

{\gt{Comparison}} constraint (R25) specified as \{$signIn\ {\gt{delayFor}}\ 250\ {\gt{on}}\ ms\}\ \textcolor{red}{\preceq_p}\ \
\ \\ \{signIn\ {\gt{delayFor}}\ \sum WCET\ {\gt{on}}\ ms\}$ can be model using the \emph{DelayFor} STA. Two new clocks $CU$, $com$ are generated: $CU \triangleq \{signIn\ {\gt{delayFor}}\ 250\ {\gt{on}}\ ms\}$, $com \triangleq \{signIn\ {\gt{delayFor}}\ \sum WCET\ {\gt{on}}\ ms\}$, where $\sum WCET$ represents the sum of worst case execution time of {\gt{Controller}} and {\gt{VehicleDynamics}} \fp s. Therefore, R25 can be expressed as the query: $Pr[\leqslant bound]([\ ]\ (ex_{con} == wcet_{con} \wedge ex_{vd} == wcet_{vd})\ \implies\ (h_{con}\ \geqslant h_{CU})\ \geqslant\ P$, where $ex_{con} == wcet_{con} \wedge ex_{vd} == wcet_{vd}$ restricts that when the execution is the worst case (i.e., the execution time is the longest), the {\gt{probabilistic causality}} relation between \emph{con} and \emph{CU} should be guaranteed.

\vspace{0.05in}
\noindent\textbf{Probabilistic Precedence} is utilized to specify \ed\ {\gt{End-to-End}} timing constraint (R17). It states that the time duration between the \emph{source} event \emph{signIn} (input signal on the \emph{signType} port of {\gt{Controller}}) and the \emph{target} event \emph{spOut} (output signal on the \emph{speed} port of {\gt{VehicleDynamic}})  must be within a time bound of [150, 250], and that is specified as \smc\ quires (56) and (57):
\begin{equation}
\{signIn\ {\gt{delayFor}}\ 150\ {\gt{on}} \ ms\}\ \textcolor{red}{\prec_p}\ spOut
\label{e2e150}
\end{equation}
\begin{equation}
spOut\ \textcolor{red}{\prec_p}\ \{signIn\ {\gt{delayFor}}\ 250\ {\gt{on}}\ ms\}
\label{e2e250}
\end{equation}

Two clocks, $lower$ and $upper$, are defined by delaying $signIn$ for 150 and 250 ticks of $ms$ respectively:
$lower \triangleq \{signIn\ {\gt{delayFor}}\ 150\ {\gt{on}}\ ms\}$, and $upper \triangleq \{signIn\ {\gt{delayFor}}\ 250\ {\gt{on}}\ ms\}$. The corresponding \smc\ models of $lower$ and $upper$ are constructed based on the {\gt{delayFor}} STA (shown in Fig. \ref{fig:CCSL_smc}). Finally, the R17 specified in Pr\ccsl\ is expressed as \smc\ quires (3) and (4), where $h_{{\gt{lower}}}$, $h_{{\gt{upper}}}$ and $h_{{\gt{spOut}}}$ are the history of $lower$, $upper$ and $spOut$. $t_{{\gt{spOut}}}$ and $t_{{\gt{upper}}}$ represent the tick of $upper$ and $spOut$ respectively:
\begin{equation}
Pr[ \leqslant bound]([\ ] h_{{\gt{lower}}} \geqslant h_{{\gt{spOut}}} \wedge ((h_{{\gt{lower}}}==h_{{\gt{spOut}}})\implies\ t_{{\gt{spOut}}} ==0)) \geqslant P
\label{e2equery1}
\end{equation}
\begin{equation}
Pr[ \leqslant bound]([\ ] h_{{\gt{spOut}}} \geqslant h_{{\gt{upper}}} \wedge ((h_{{\gt{spOut}}}==h_{{\gt{upper}}})\implies\ t_{{\gt{upper}}} ==0)) \geqslant P
\label{e2equery2}
\end{equation}

\noindent In similar, \ed\ {\gt{Sporadic}} timing constraint (R9) specified in {\gt{probabilistic precedence}}  can be translated into \emph{Hypothesis Testing} query $Pr[ \leqslant bound]([\ ] ho \geqslant hv \wedge ((hv==ho)\implies\ t_{{\gt{va}}} ==0)) \geqslant P$, where $ho$ represents the history of the clock/event that obstacle occurs, and $t_{va}$ and $hv$ indicates the ticks and history of the clock that the vehicle starts to move.

\vspace{0.05in}
In the case of properties specified in either {\gt{probabilistic subclock}} or {\gt{probabilistic coincidence}}, such properties can not be directly expressed as \smc\ queries. Therefore, we construct an observer STA that captures the semantics of standard {\gt{subclock}} and {\gt{coincidence}} \emph{relations}. The observer STA are composed to the system STA, namely a network STA NSTA, in parallel. Then, the probabilistic analysis is performed over the NSTA  which enables us to verify the \ed\ timing constraints specified in {\gt{probabilistic subclock}} and {\gt{probabilistic coincidence}} of the entire system using \smc. Further details are given below.

\begin{figure}[htbp]
\centering
  \subfigure[Subclock]{
  \includegraphics[width=2in]{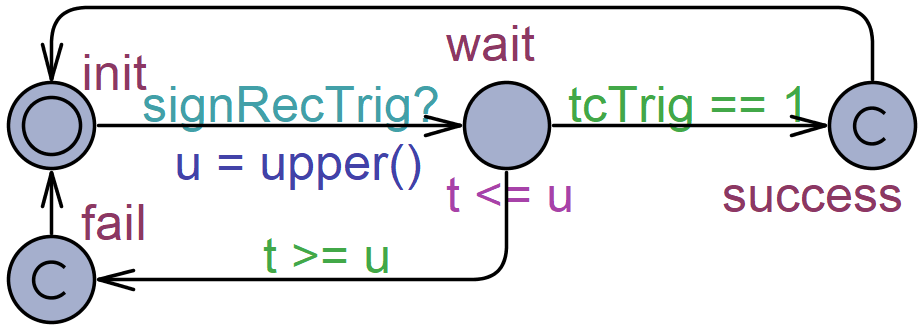}
  \label{fig:subclock}}
  \subfigure[Coincidence]{
  \includegraphics[width=2.4in]{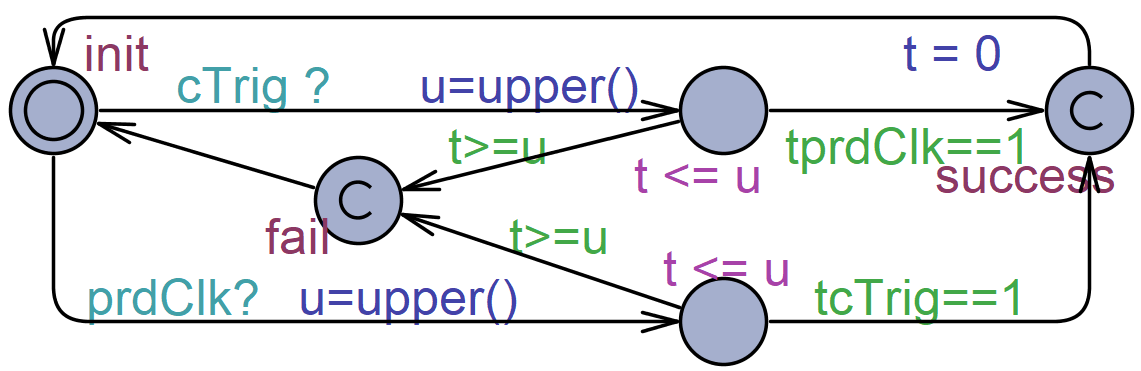}
  \label{fig:coin}}
  \caption{Observer STA of {\gt{Subclock}} and {\gt{Coincidence}}}
\label{fig:PRCCSL_smc}
\end{figure}

\vspace{0.05in}
\noindent\textbf{Probabilistic Subclock} is employed to specify \ed\ {\gt{Periodic}} timing constraint, given as $signRecTrig$ $\textcolor{red}{\subseteq_p}$ $cTrig$ (Spec. R2 in Fig.1). The standard {\gt{subclock}} \emph{relation} states that \emph{superclock} must tick at the same step where \emph{subclock} ticks. Its corresponding STA is shown in Fig. \ref{fig:PRCCSL_smc}.(a).
When $signRevTrig$ ticks ($signRecTrig?$), the STA transits to the $wait$ location and detects the occurrence of $cTrig$ until the time point of the subsequent step ($u$).
If $cTrig$ occurs prior to the next step ($tcTrig==1$), the STA moves to the $success$ location, i.e., the {\gt{subclock}} \emph{relation} is satisfied at the current step. Otherwise, it  transits to the $fail$ location. R2 specified in {\gt{probabilistic subclock}} is expressed as: $Pr[bound] ([\ ] \neg\ Subclock.fail) \geqslant P$. \smc\ analyzes if the $fail$ location is never reachable from the system NSTA, and whether the probability of R2 being satisfied is greater than or equal to $P$.

\vspace{0.05in}
\noindent\textbf{Probabilistic Coincidence} is adapted to specify \ed\ {\gt{Periodic}} timing constraint, given as $cTrig$ $\textcolor{red}{\equiv_p}$ $\{{\gt{periodicOn}}$ $ms$ ${\gt{period}}$ $50\}$ (Spec. R1 in Fig.1). To express R1 in \smc,
first, a periodic clock \emph{prdClk} ticking every $50^{th}$ tick of $ms$ is defined: $prdClk$ $\triangleq$ ${\gt{periodicOn}}$ $ms$ ${\gt{period}}$ $50$. The corresponding \smc\ model of
$prdClk$ is generated based on the {\gt{periodicOn}} STA shown in Fig. \ref{fig:CCSL_smc}.(a) by setting $q$ as 50.
Then, we check if $cTrig$ and $prdClk$ are coincident by employing the {\gt{coincidence}} STA shown in Fig. \ref{fig:PRCCSL_smc}.(b).  When $cTrig$ ($prdClk$) ticks via $cTrig?$ ($prdClk?$), the STA checks if the other clock, $prdClk$ ($cTrig$), ticks prior to the next step, i.e., whether $tprdClk==1$ ($tcTrig==1$) holds or not when $t \leqslant u$. The STA then transits to either the $success$ or $fail$ location based on the judgement. R1 specified in {\gt{probabilistic coincidence}} is expressed as: $Pr[bound] ([\ ] \neg\ Coincidence.fail) \geqslant P$. \smc\ analyzes if the probability of R1 being satisfied is greater than or equal to $P$.

\chapter{Modeling the Behaviors of AV and its Environment in UPPAAL-SMC}
\label{sec:beha}
\begin{figure}[htbp]
\centering
  \subfigure[{\gt{Obstacle}}]{
  \includegraphics[width=3.6in]{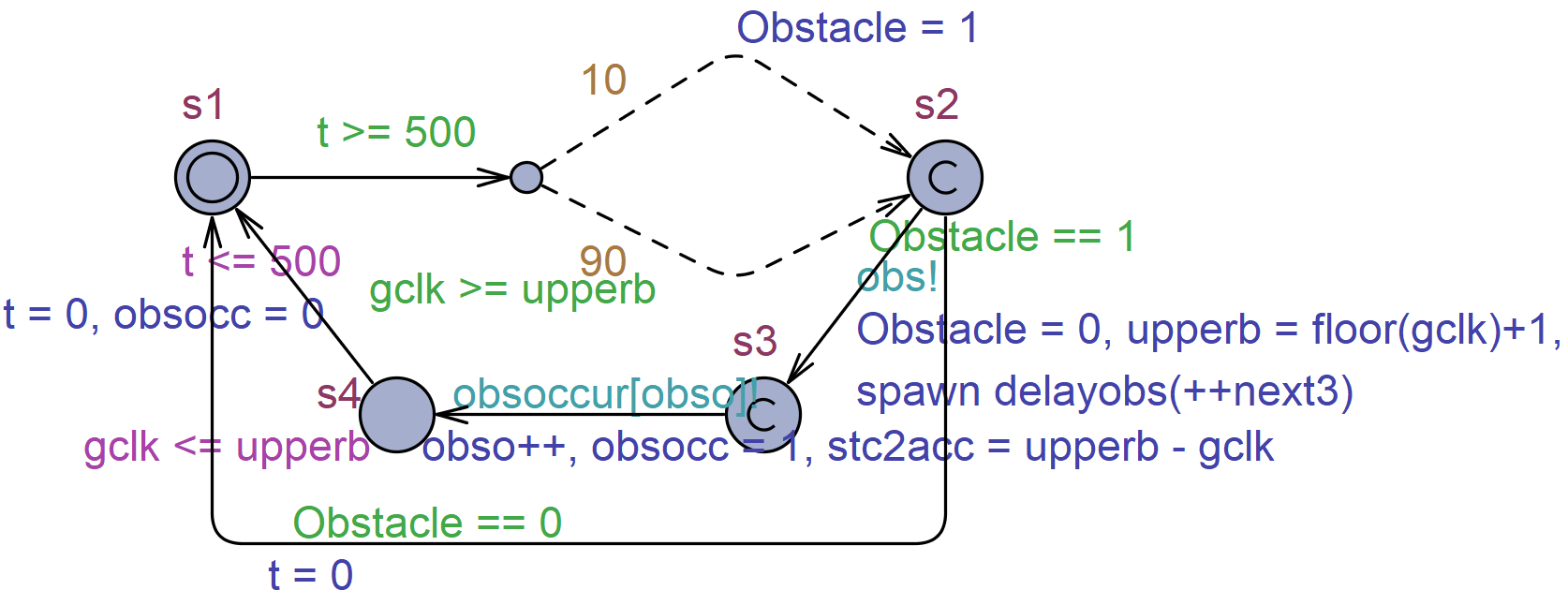}
  \label{fig:obs}}
  \subfigure[{\gt{Speed}}]{
  \includegraphics[width=2.3in]{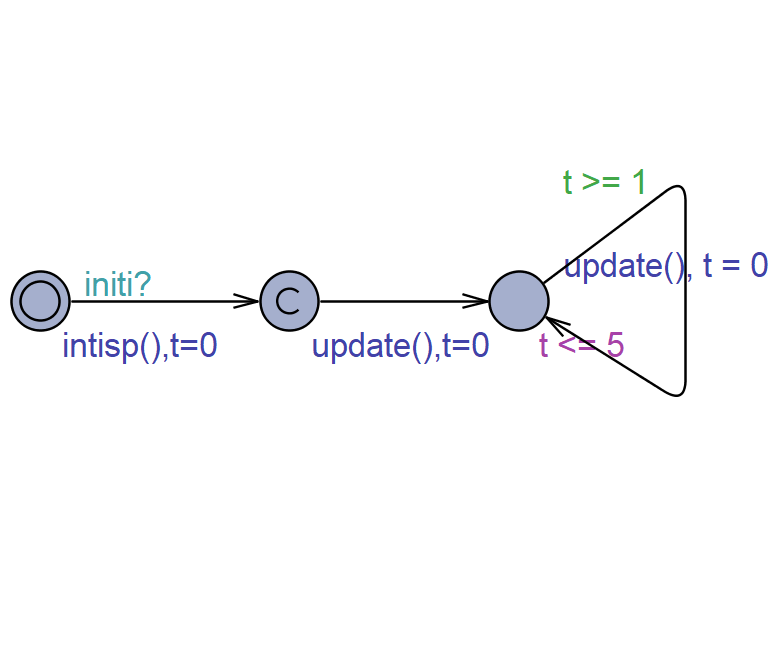}
  \label{fig:speed}}
  \subfigure[{\gt{TrafficSign}}]{
  \includegraphics[width=3.8in]{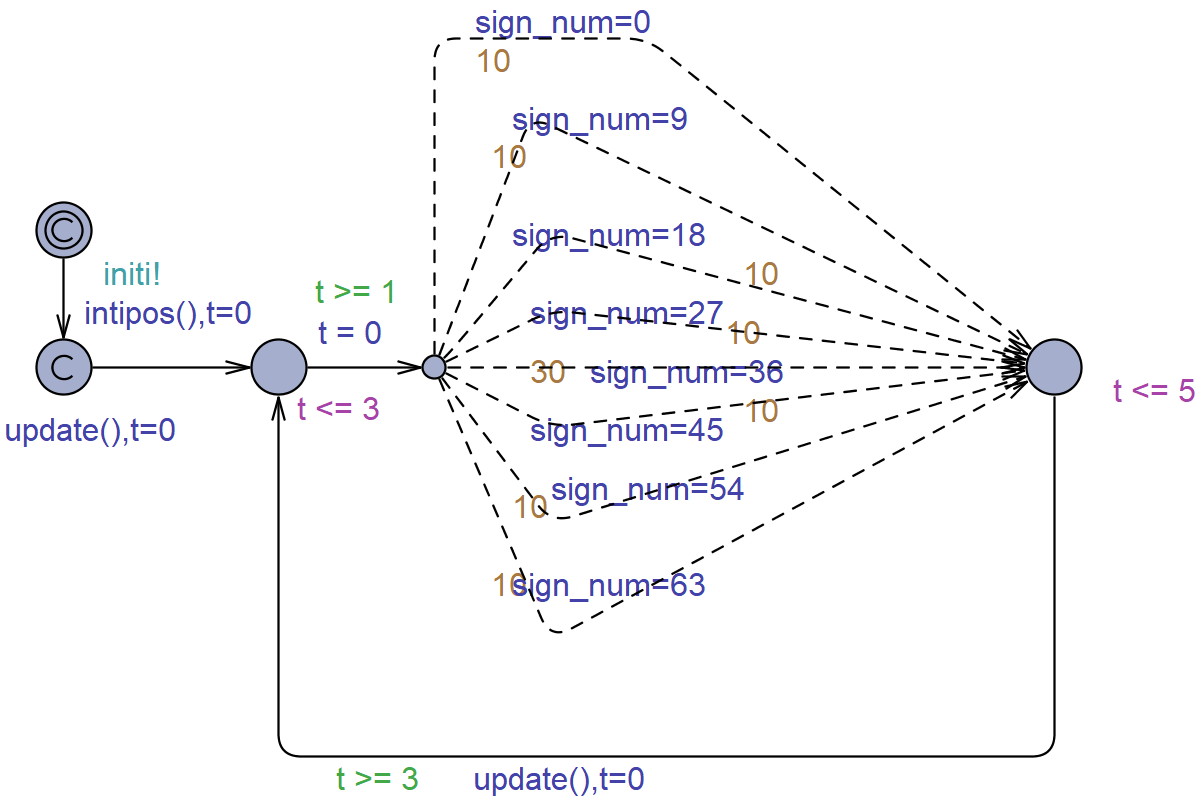}
  \label{fig:sign}}
  \caption{STAs of \fp s in \gt{Environment} \ft}
\label{fig:env}
\end{figure}

\begin{figure}[htbp]
\centering
  \subfigure[{\gt{Camera}}]{
  \includegraphics[width=3.5in]{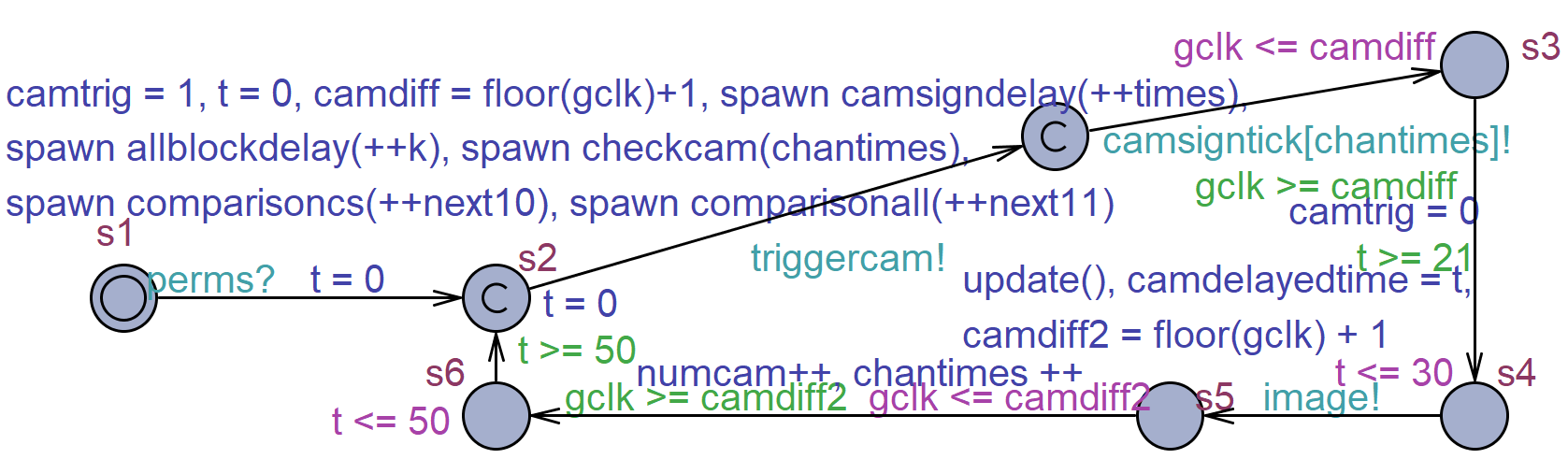}
  \label{fig:cam}}
  \subfigure[{\gt{SignRecognition}}]{
  \includegraphics[width=3.5in]{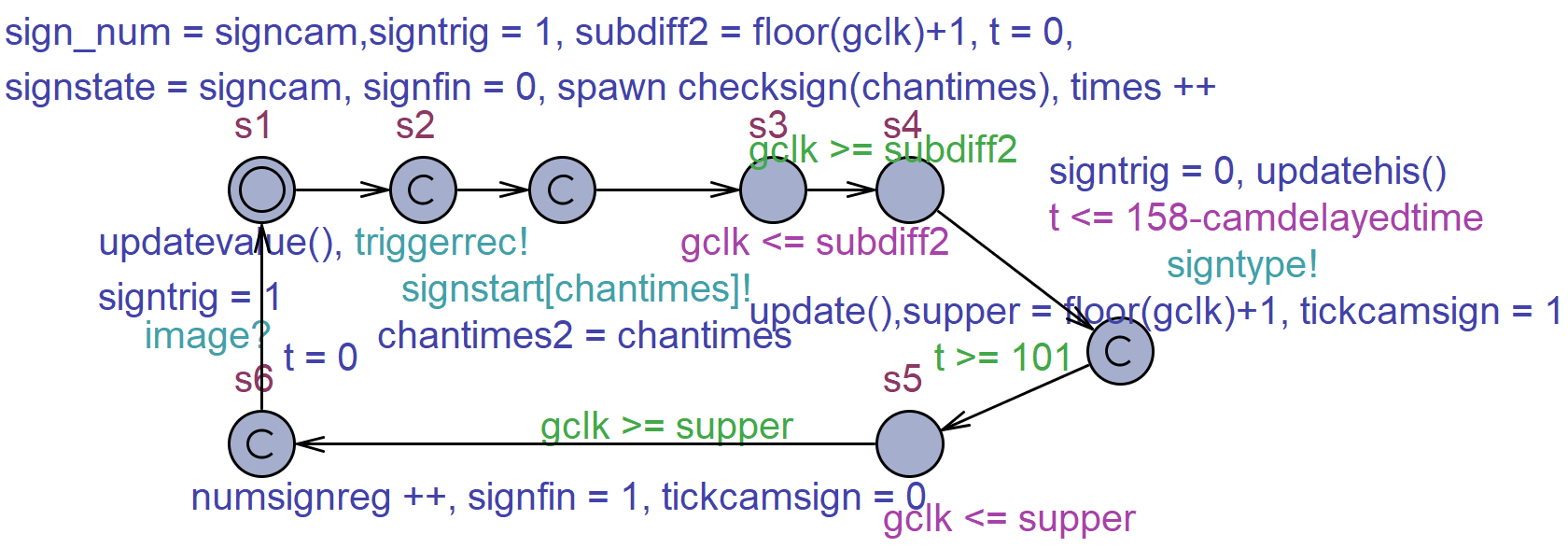}
  \label{fig:signreg}}
  \subfigure[{\gt{Controller}}]{
  \includegraphics[width=3.5in]{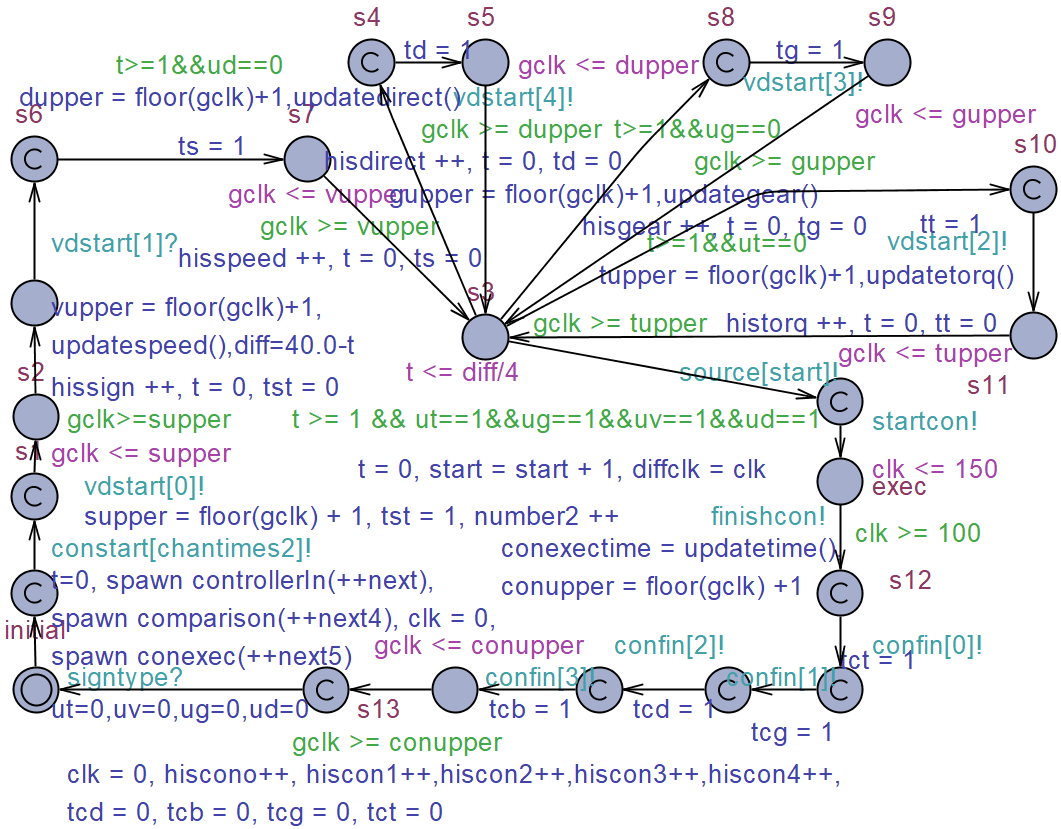}
  \label{fig:con}}
  \subfigure[{\gt{VehicleDynamic}}]{
  \includegraphics[width=3.7in]{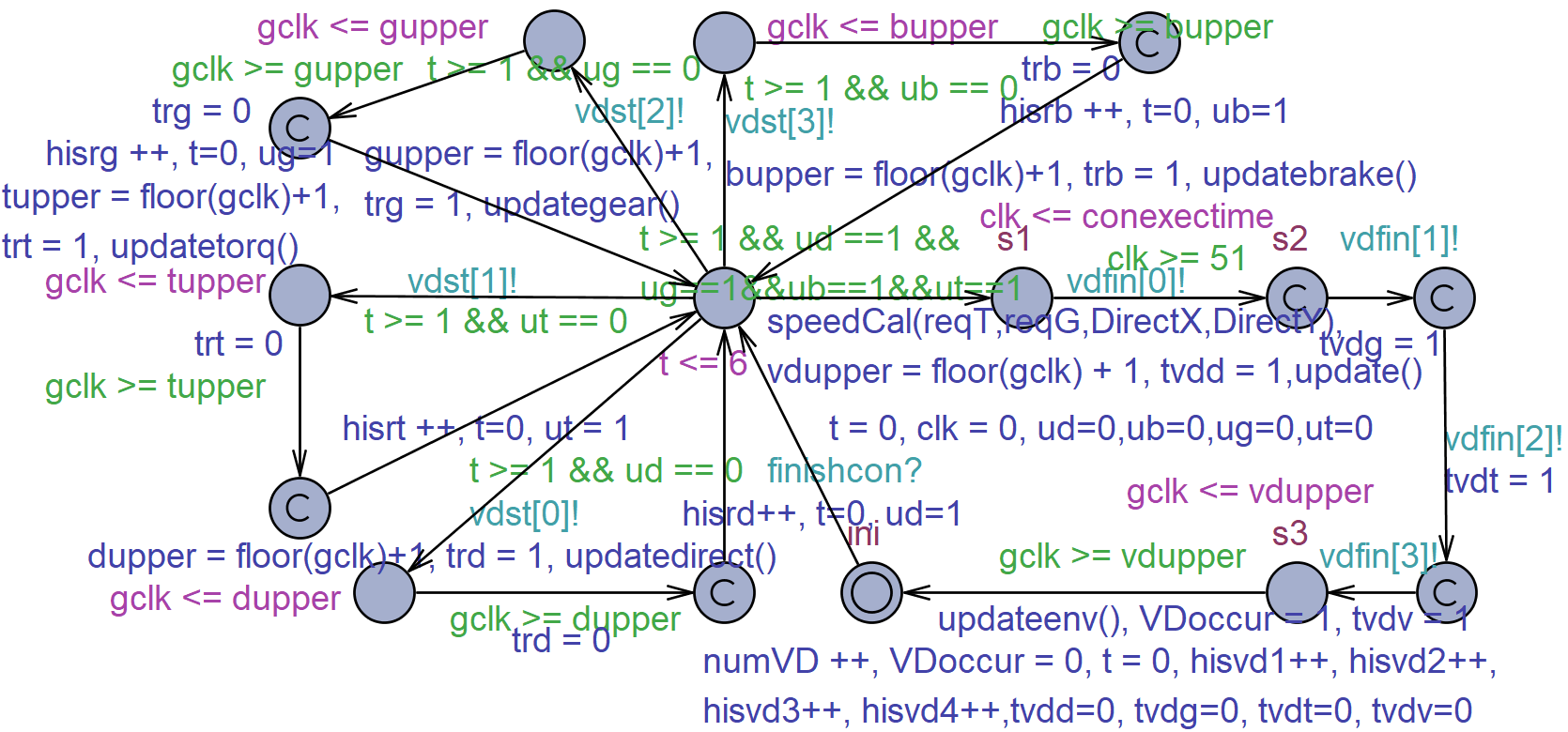}
  \label{fig:vd}}
  \caption{Modeling system behaviors in \smc}
\label{fig:sys}
\end{figure}

To capture the behaviours of the AV system and the stochastic behaviours of its environments, e.g., random traffic signs, each \fp\ in Fig. \ref{fig:AV_model} is modeled as an STA in \smc. The random traffic sign in the environment is recognised by AV. The speed of the AV is influenced by the condition of the road. Obstacles on the road occurs randomly. To model these stochastic behaviours, we model the three \fp s in the {\gt{Environment}} \ft\ into three STAs, which are presented in Fig. \ref{fig:env}. In {\gt{TrafficSign}} (Fig. \ref{fig:sys}.(b)) STA, $sign\_{num}$ represents the random traffic sign type, which is generated every 4ms to 8ms. To represent the integration of the AV system and the environment, the speed of AV is equal to the speed of in the environment, the {\gt{Speed}} (shown in Fig. \ref{fig:env}.(c)) STA updates the speed of the vehicle in the environment from the by activating the execution of $update()$ function periodically. {\gt{Obstacle}} STA generates a signal randomly based on probability distribution to represent random obstacles.

\begin{figure}[htbp]
\centering
  \subfigure[{\gt{speed}}]{
  \includegraphics[width=2.5in]{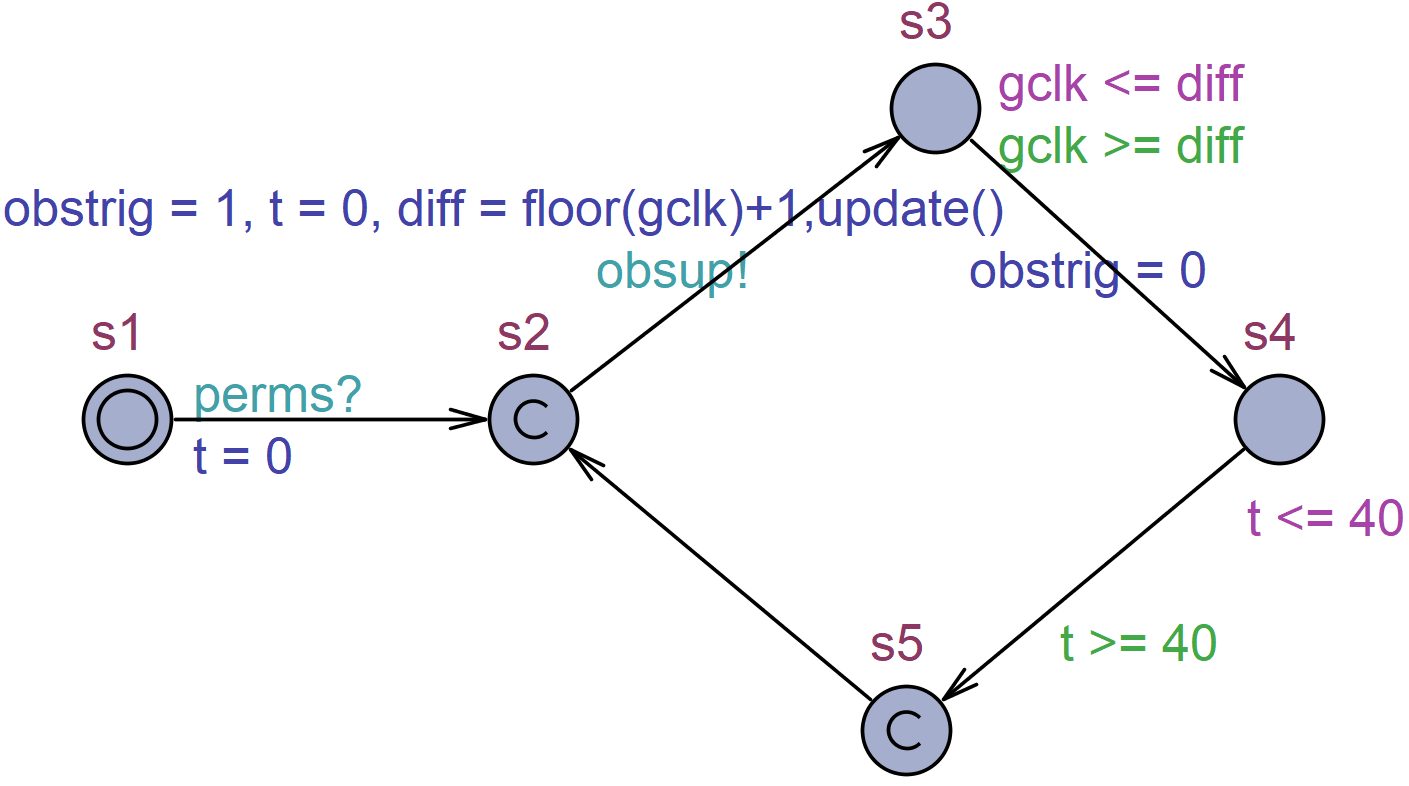}
  \label{fig:speedup}}
  \subfigure[{\gt{obstacle}}]{
  \includegraphics[width=2.5in]{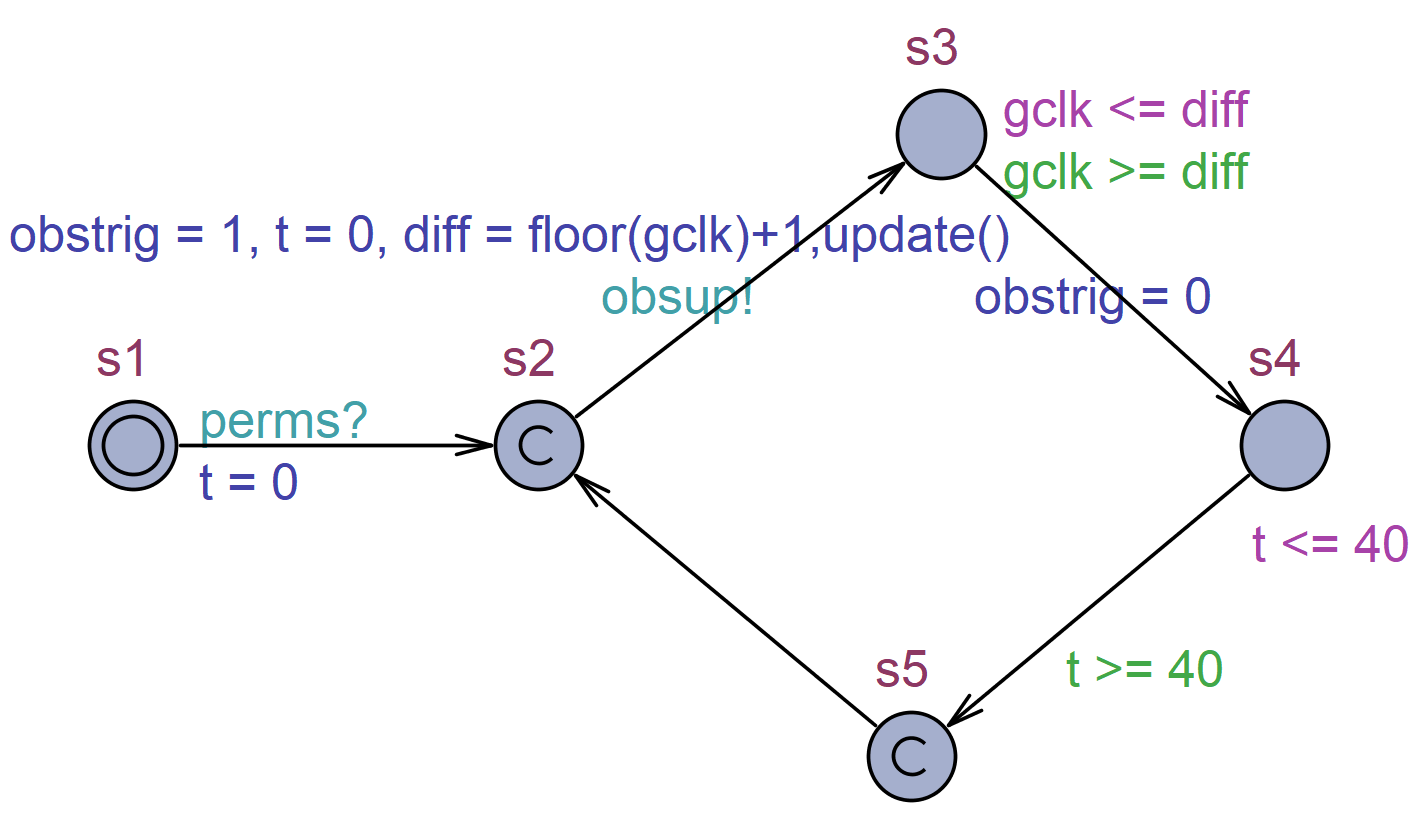}
  \label{fig:obsup}}
  \caption{Periodically triggered ports {\gt{speed}} and {\gt{obstacle}}}
\label{fig:perup}
\end{figure}

\begin{figure}[htbp]
\centering
  \subfigure[Top View]{
  \includegraphics[width=3.0in]{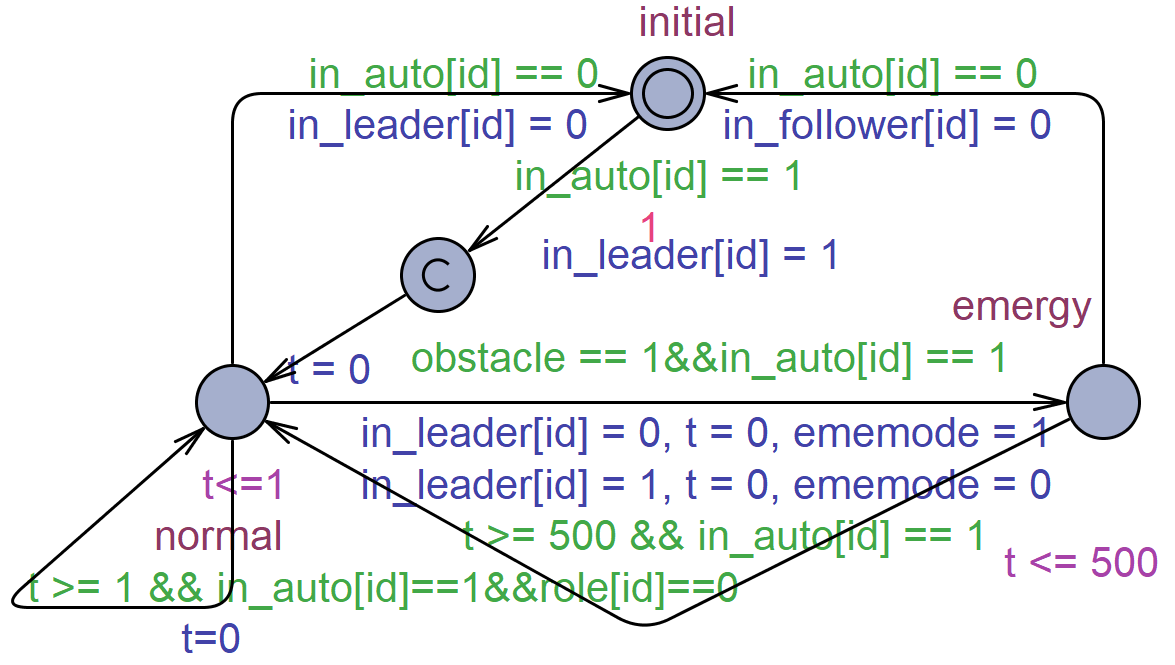}
  \label{fig:topview}}
  \subfigure[{\gt{Emergency}}]{
  \includegraphics[width=2.0in]{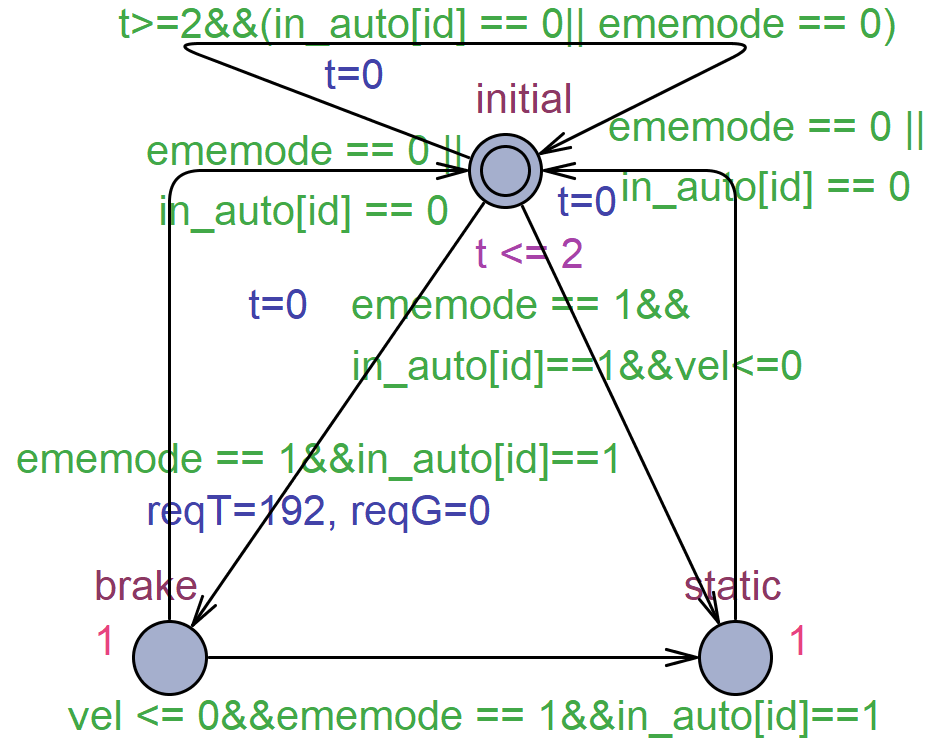}
  \label{fig:eme}}
  \subfigure[{\gt{Normal}}]{
  \includegraphics[width=4.8in]{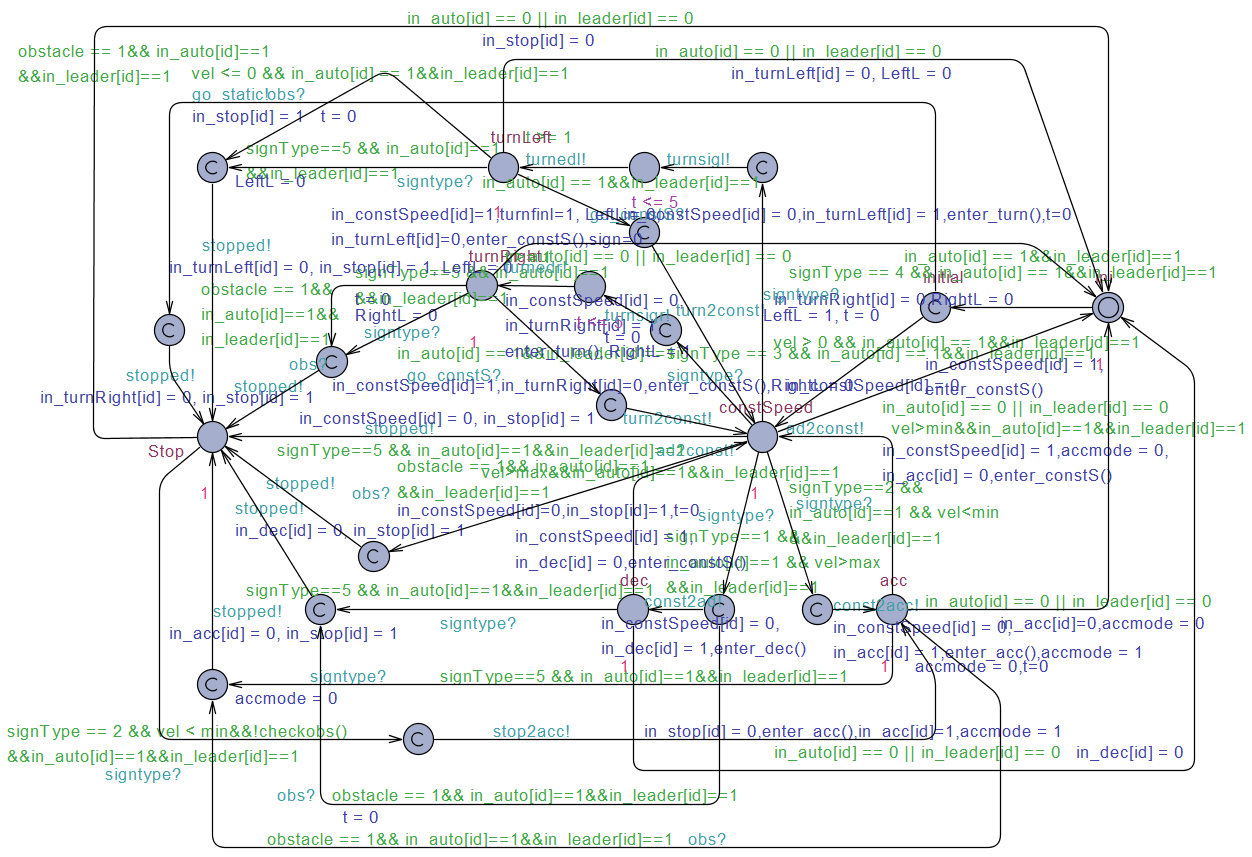}
  \label{fig:normal}}
  \caption{Internal behaviours of {\gt{Controller}} in UPPAAL-SMC}
\label{fig:controller}
\end{figure}

\begin{figure}[htbp]
\centering
  \subfigure[{\gt{Acceleration}}]{
  \includegraphics[width=2.0in]{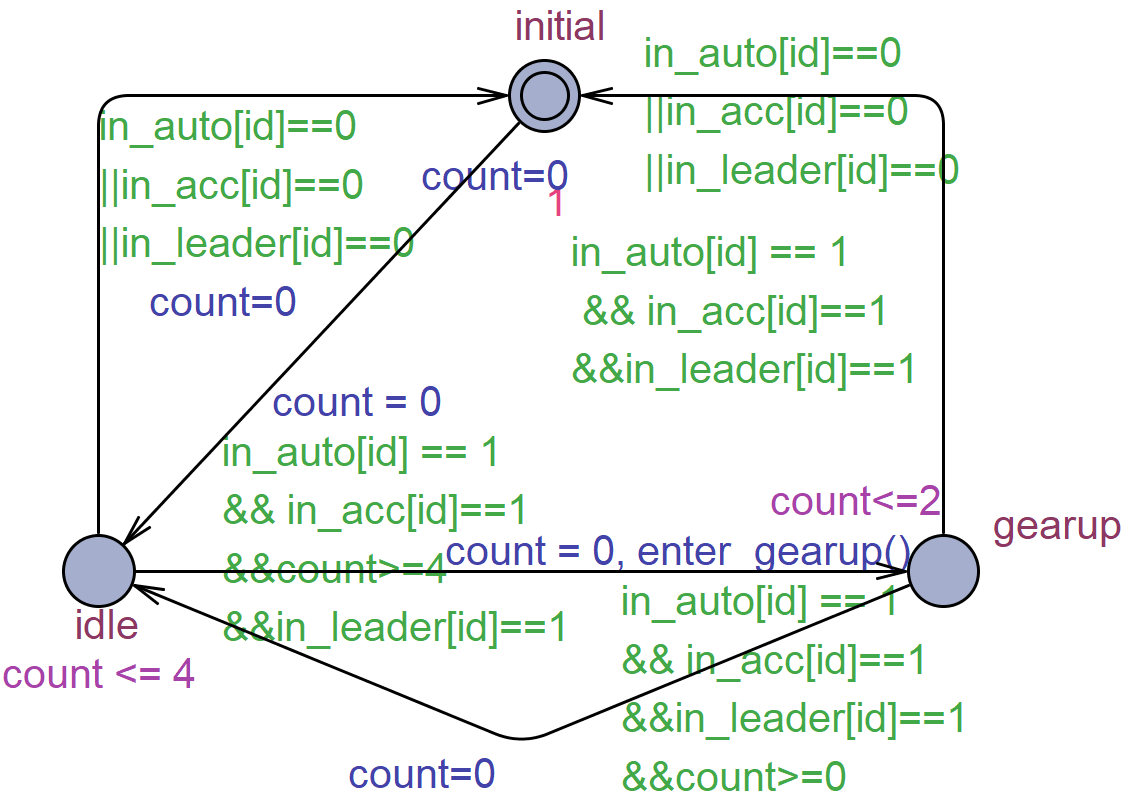}
  \label{fig:acc}}
  \subfigure[{\gt{Deceleration}}]{
  \includegraphics[width=1.8in]{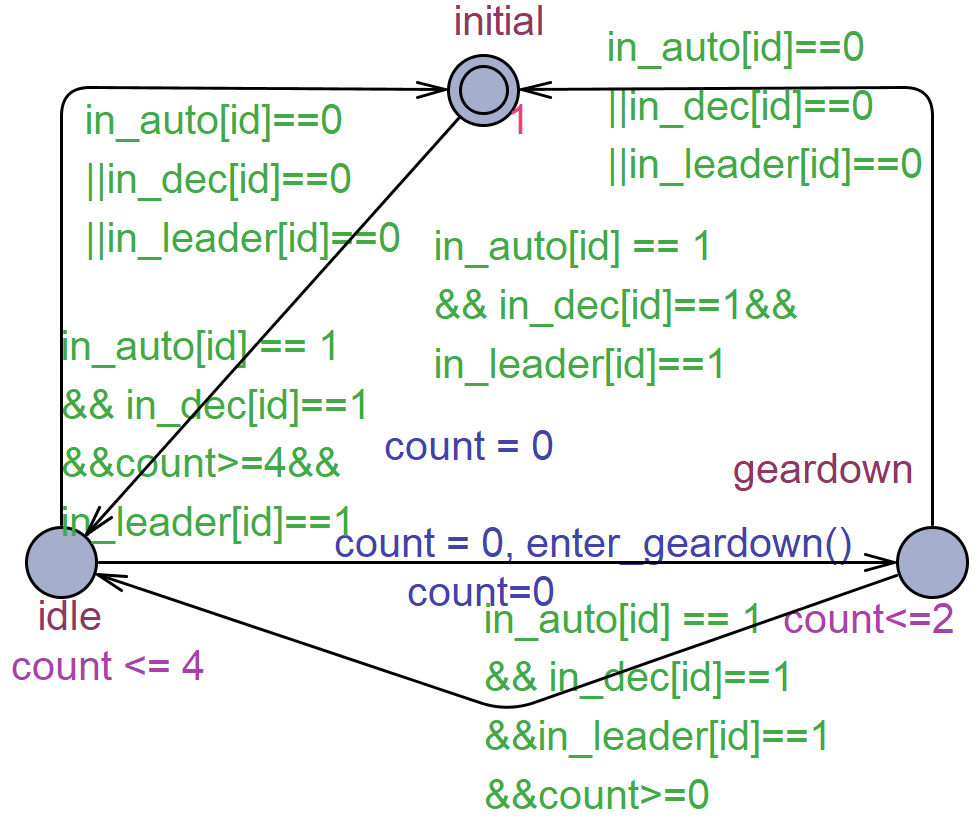}
  \label{fig:dec}}
  \subfigure[{\gt{Stop}}]{
  \includegraphics[width=3.0in]{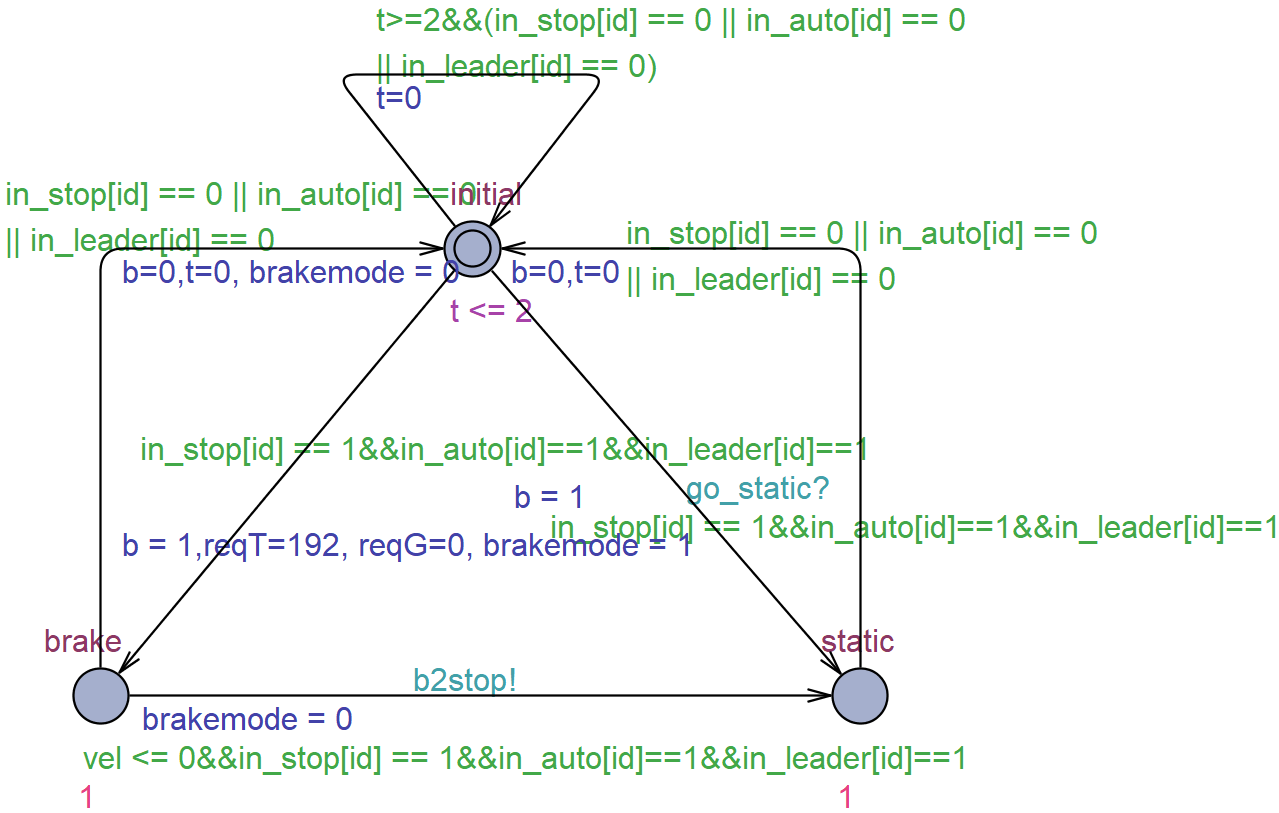}
  \label{fig:stop}}
  \subfigure[{\gt{TurnLeft}}]{
  \includegraphics[width=4.2in]{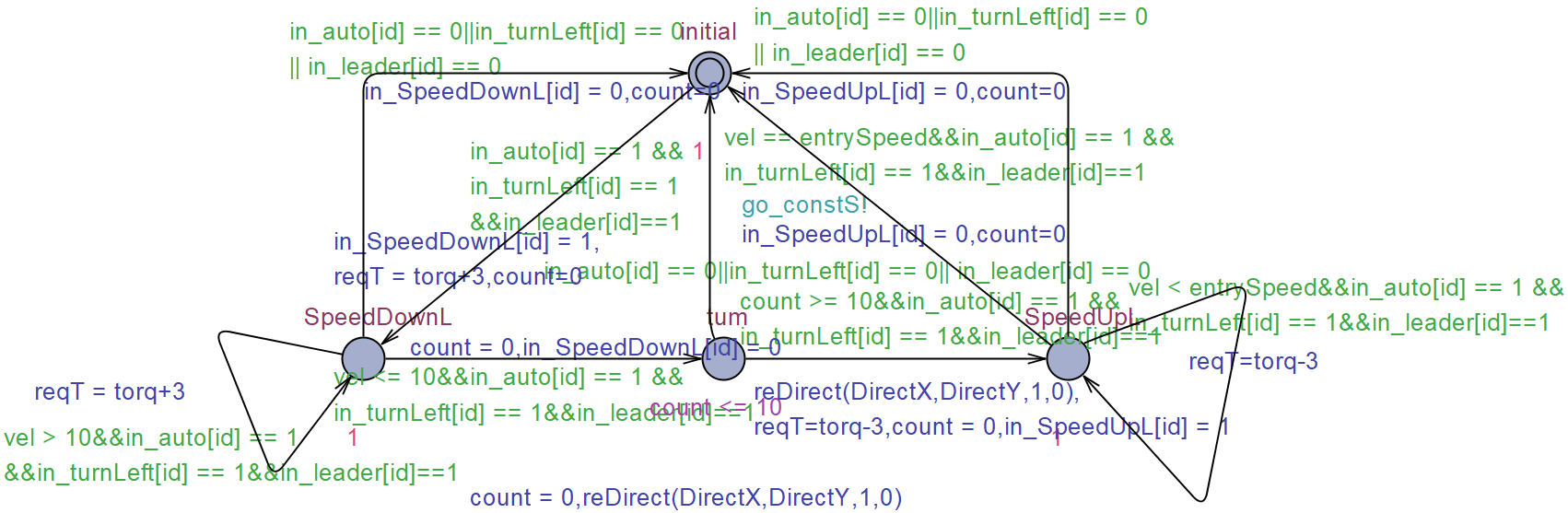}
  \label{fig:tl}}
  \subfigure[{\gt{TurnRight}}]{
  \includegraphics[width=4.2in]{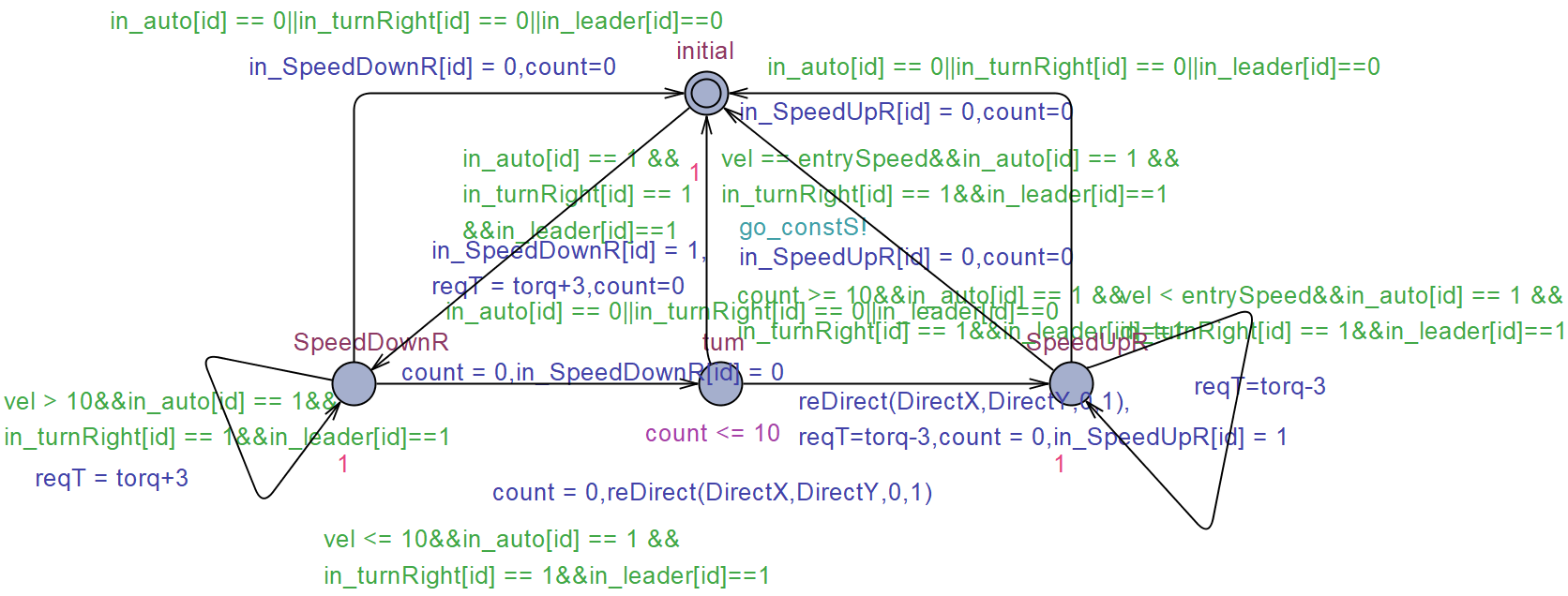}
  \label{fig:tr}}
  \caption{Representation of substates of \gt{Normal} state in UPPAAL-SMC}
\label{fig:unormal}
\end{figure}

\begin{figure}[htbp]
\centering
  \subfigure[{\gt{Periodic}} STA with period 40ms]{
  \includegraphics[width=2.8in]{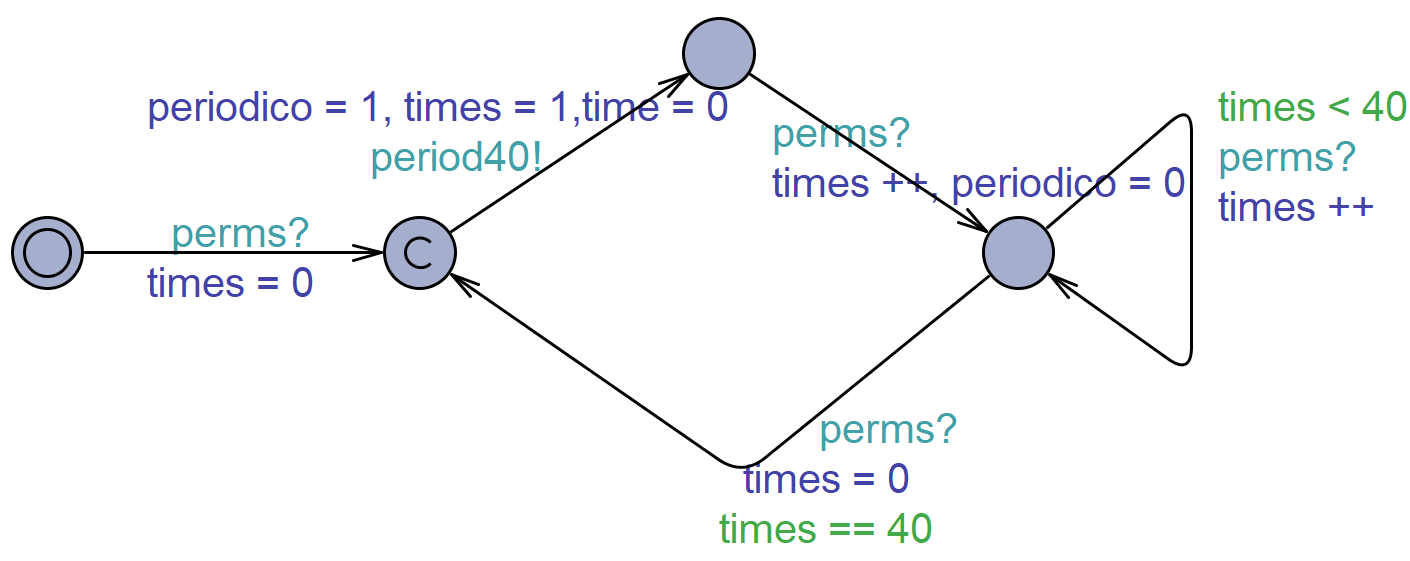}
  \label{fig:p3}}
  \subfigure[{\gt{Coincidence}} STA]{
  \includegraphics[width=2.4in]{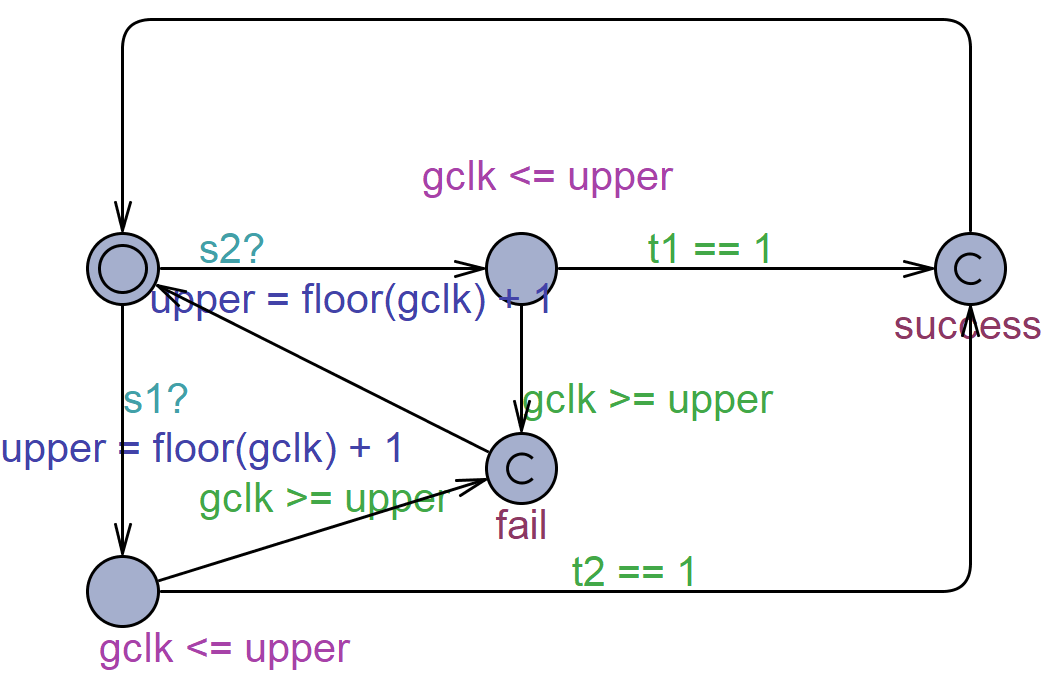}
  \label{fig:c3}}
  \caption{STAs utilized to verify R3}
\label{fig:R3}
\end{figure}

The system model of AV is represented as the STAs shown in Fig. \ref{fig:sys}. {\gt{Camera}} STA is  triggered periodically (Fig. \ref{fig:sys}.(a)). When the execution of camera is finished, i.e., the transition from $s4$ to $s5$ is taken, the $update()$ function is triggered and the value of $sign\_num$ is assigned to $signType$. Since the input ports ({\gt{speed}} and {\gt{obstacle}}) of {\gt{Controller}} are triggered periodically, the AV system obtains the speed of the vehicle and the road information by executing the $update()$ periodically (Fig. \ref{fig:perup}).

The internal behaviours of {\gt{Controller}} \fp\ is captured in Fig. \ref{fig:controller}. When the vehicle is in the ``normal'' mode (Fig. \ref{fig:controller}.(c)) and it encounters an obstacle, the ``emergency stop'' mode will be activated (Fig. \ref{fig:controller}.(b)) and the vehicle begins to stop. In ``normal'' mode, the vehicle adjusts its movement according to the traffic signs, e.g., when it detects a turn left sign, it will turn left (Fig. \ref{fig:unormal}.(d)). The {\gt{Controller}} then sends out requests for {\gt{VehicleDynamic}} to change the direction or the speed of the four wheels.

To verify R1 to R31, STAs of \ccsl\ $expressions$ {\gt{periodicOn}}, {\gt{infimum}}, {\gt{supremum}} and {\gt{delayFor}} and STAs of Pr\ccsl\ $relations$ {\gt{coincidence}} and {\gt{subclock}} are utilized. For example, to verify R3, a {\gt{periodicOn}} STA generates a new clock $c$ with period 40 (Fig. \ref{fig:R3}.(a)). When $c$ ticks, the $periodico$ will be assigned to 1. The {\gt{probabilistic coincidence}} \emph{relation} between  $c$ and the triggering of the {\gt{obstacle}} port should hold. When the input port is triggered, $obstrig$ will become 1 in Fig. \ref{fig:perup}.(b). {\gt{Coincidence}} STA (Fig. \ref{fig:R3}.(b)) is employed for checking the {\gt{coincidence}} $relation$ between $c$ and $obstacle$.
\chapter{Experiments: Verification \& Validation}
\label{sec:experiment}
\begin{table*}[htbp]
 \scriptsize
  \centering
   \renewcommand\arraystretch{1.3}
       \caption{Verification Results in \smc}
\begin{tabular}[htbp]{|c|c|c|p{220pt}|c|c|c|c|}
    \hline
    Type & R.ID & Q & Expression & Result & Time & Mem & CPU \\
    \hline
    \multirow{7}{*}{Periodic} & \multirow{5}{*}{R1} &HT & Pr[$\leqslant$3000]([ ] $\neg Coin.fail$)$\geqslant$0.95  & valid & 48.7 & 32.7 & 31.3 \\\cline{3-8}
    \multirow{7}{*}{} & \multirow{5}{*}{} & PE & Pr[$\leqslant$3000]([ ] $\neg Coin.fail$)  & [0.902, 1] & 12.6 & 35.6 & 29.8 \\\cline{3-8}
    \multirow{7}{*}{} & \multirow{5}{*}{} & EV & E[$\leqslant$3000; 500]([ ] $max:cam.t$) & 50$\pm$0 & 83.3 & 33.3 & 31.7 \\\cline{3-8}
    \multirow{7}{*}{} & \multirow{5}{*}{} & SI& simulate 500 [$\leqslant$3000]($camtrig,\ p1trig$)  & valid & 80.9 & 32.9 & 32.5\\\cline{2-8}
    \multirow{7}{*}{} & \multirow{5}{*}{R2}  & HT & Pr[$\leqslant$3000]([ ] $\neg Sub.fail$)$\geqslant$0.95  & valid & 48.9 & 32.9 & 29.3\\\cline{3-8}
    \multirow{7}{*}{} & \multirow{5}{*}{}& PE & Pr[$\leqslant$3000]([ ] $\neg Sub.fail$)  & [0.902, 1] & 12.3 & 35.5 & 30.4\\\cline{3-8}
    \multirow{7}{*}{} & \multirow{5}{*}{}& EV & E[$\leqslant$3000; 500]([ ] $max:sf.t$)  & 200$\pm$0 & 80.6 & 32.5 & 32.2\\\cline{3-8}
    \multirow{7}{*}{} & \multirow{5}{*}{}& SI& simulate 500 [$\leqslant$3000]($strig,\ p2trig$)  & valid & 85.5 & 33.1 & 32.3\\\cline{2-8}
    \multirow{7}{*}{}& \multirow{2}{*}{R3}  &HT & Pr[$\leqslant$3000]([ ] $\neg Coin_{obs}.fail$)$\geqslant$0.95  & valid & 57.6 & 40.5 & 34.6 \\\cline{3-8}
    \multirow{7}{*}{} & \multirow{2}{*}{} & PE & Pr[$\leqslant$3000]([ ] $\neg Coin_{obs}.fail$) & [0.902, 1] & 13.8 & 40.4 & 31.1 \\\cline{2-8}
    \multirow{7}{*}{}& \multirow{2}{*}{R4} & HT & Pr[$\leqslant$3000]([ ] $\neg Coin_{sp}.fail$)$\geqslant$0.95 & valid & 56.7 & 40.4 & 32.4 \\\cline{3-8}
    \multirow{7}{*}{} & \multirow{2}{*}{} & PE & Pr[$\leqslant$3000]([ ] $\neg Coin_{sp}.fail$) & [0.902, 1] & 13.6 & 35.9 & 34.0 \\
   \hline
   \multirow{7}{*}{Execution} & \multirow{5}{*}{R5} & HT  & Pr[$\leqslant$3000]([ ] $h_{{\gt{SU}}} \leqslant h_{{\gt{S}}}$) $\geqslant$ 0.95  & valid & 76.5 & 40.4 & 32.3\\\cline{3-8}
    \multirow{7}{*}{} & \multirow{5}{*}{}& PE & Pr[$\leqslant$3000]([ ] $h_{{\gt{SU}}} \leqslant h_{{\gt{S}}}$)  & [0.902, 1] & 18.1 & 40.3 & 30.8\\\cline{3-8}
    \multirow{7}{*}{} & \multirow{5}{*}{}& HT  & Pr[$\leqslant$3000]([ ] $h_{{\gt{S}}} \leqslant h_{{\gt{SL}}}$) $\geqslant$ 0.95  & valid & 77.6 & 37.7 & 31.7\\\cline{3-8}
    \multirow{7}{*}{} & \multirow{5}{*}{}& PE & Pr[$\leqslant$3000]([ ] $h_{{\gt{S}}} \leqslant h_{{\gt{SL}}}$)  & [0.902, 1] & 16.5 & 40.0 & 31.5\\\cline{3-8}
    \multirow{7}{*}{} & \multirow{5}{*}{}& PC & Pr[$\leqslant$3000] ([ ] $SR.exec \implies$ ($SR.t \geqslant 100 \wedge SR.t \leqslant 125$)) $\geqslant$ Pr[$\leqslant$3000] ([ ] $SR.exec \implies$ ($SR.t \geqslant 125 \wedge SR.t \leqslant 150$)) & $\geqslant$1.1 & 8.3& 31.7 & 32.3\\\cline{3-8}
    \multirow{7}{*}{} & \multirow{5}{*}{}& EV & E[$\leqslant$3000; 500]([ ] $max:checkexe.t$)  & 147.2$\pm$0.7 & 85.8 & 32.4 & 36.0 \\\cline{3-8}
    \multirow{7}{*}{} & \multirow{5}{*}{}& SI & simulate 500 [$\leqslant$3000]($h_{{\gt{SU}}}, h_{{\gt{S}}}, h_{{\gt{SL}}}$)  & valid & 89.5 & 34.0 & 34.0\\\cline{2-8}
    \multirow{7}{*}{}& \multirow{5}{*}{R6}  &HT  & Pr[$\leqslant$3000]([ ] $h_{{\gt{camU}}} \leqslant h_{{\gt{C}}}$) $\geqslant$ 0.95  & valid & 42.8 & 37.9 & 33.2\\\cline{3-8}
    \multirow{7}{*}{} & \multirow{5}{*}{}& PE & Pr[$\leqslant$3000]([ ] $h_{{\gt{camU}}} \leqslant h_{{\gt{C}}}$)  & [0.902, 1] & 10.8 & 37.9 & 30.8\\\cline{3-8}
    \multirow{7}{*}{} & \multirow{5}{*}{}& HT  & Pr[$\leqslant$3000]([ ] $h_{{\gt{C}}} \leqslant h_{{\gt{camL}}}$) $\geqslant$ 0.95  & valid & 38.1 & 34.4 & 32.3\\\cline{3-8}
    \multirow{7}{*}{} & \multirow{5}{*}{}& PE & Pr[$\leqslant$3000]([ ] $h_{{\gt{C}}} \leqslant h_{{\gt{camL}}}$)  & [0.902, 1] & 9.9 & 37.9 & 31.2\\\cline{3-8}
    \multirow{7}{*}{} & \multirow{5}{*}{}& PC & Pr[$\leqslant$3000] ([ ] $cam.exec \implies$ ($cam.t \geqslant 20 \wedge cam.t \leqslant 25$)) $\geqslant$ Pr[$\leqslant$3000] ([ ] $cam.exec \implies$ ($cam.t \geqslant 25 \wedge cam.t \leqslant 30$)) & $\geqslant$1.1 & 4s & 34.0 & 30.9\\\cline{2-8}
    \multirow{7}{*}{}& \multirow{4}{*}{R7}  &HT & Pr[$\leqslant$3000]([ ] $h_{{\gt{conU}}} \leqslant h_{{\gt{Con}}}$)$\geqslant$0.95  & valid & 45.6 & 38.2 & 33.7 \\\cline{3-8}
    \multirow{7}{*}{} & \multirow{4}{*}{} &HT & Pr[$\leqslant$3000]([ ] $h_{{\gt{Con}}} \leqslant h_{{\gt{conL}}}$) $\geqslant$ 0.95  & valid & 46.3 & 38.3 & 32.6\\\cline{3-8}
    \multirow{7}{*}{} & \multirow{4}{*}{} & PC & Pr[$\leqslant$3000] ([ ] $con.exec \implies$ ($con.t \geqslant 100 \wedge con.t \leqslant 125$)) $\geqslant$ Pr[$\leqslant$3000] ([ ] $con.exec \implies$ ($con.t \geqslant 125 \wedge con.t \leqslant 150$)) & $\geqslant$1.1 & 6.9 & 34.0 & 29.3\\\cline{3-8}
    \multirow{7}{*}{} & \multirow{4}{*}{} & SI & simulate 100 [$\leqslant$3000]($h_{{\gt{conU}}}, h_{{\gt{Con}}}, h_{{\gt{conL}}}$) & valid & 33.4 & 38.8 & 34.2 \\\cline{2-8}
    \multirow{7}{*}{}& \multirow{4}{*}{R8}  &PE & Pr[$\leqslant$3000]([ ] $h_{{\gt{vdU}}} \leqslant h_{{\gt{VD}}}$) & [0.902, 1] & 14.5 & 35.8 & 35.4 \\\cline{3-8}
    \multirow{7}{*}{} & \multirow{4}{*}{} &PE & Pr[$\leqslant$3000]([ ] $h_{{\gt{VD}}} \leqslant h_{{\gt{vdL}}}$) & [0.902, 1] & 15.4 & 35.9 & 33.1\\\cline{3-8}
    \multirow{7}{*}{} & \multirow{4}{*}{} & PC & Pr[$\leqslant$3000] ([ ] $VD.exec \implies$ ($VD.t \geqslant 50 \wedge VD.t \leqslant 75$)) $\geqslant$ Pr[$\leqslant$3000] ([ ] $VD.exec \implies$ ($VD.t \geqslant 75 \wedge VD.t \leqslant 100$)) & $\geqslant$1.1 & 10.1 & 34.1 & 31.5\\\cline{3-8}
    \multirow{7}{*}{} & \multirow{4}{*}{} & SI & simulate 100 [$\leqslant$3000]($h_{{\gt{vdU}}}, h_{{\gt{VD}}}, h_{{\gt{vdL}}}$) & valid & 35.8 & 39.2 & 33.3 \\
   \hline
   \multirow{7}{*}{Sporadic} & \multirow{5}{*}{R9} & HT  & Pr[$\leqslant$3000]([ ] $hv \leqslant ho\ \wedge\ $(($hv == ho$)$\ \implies\ t_{\gt{va}} == 0$) $\geqslant$ 0.95  & valid & 3h & 33.1 & 30.0\\\cline{3-8}
    \multirow{7}{*}{} & \multirow{5}{*}{}& PE & Pr[$\leqslant$3000]([ ] $hv \leqslant ho\ \wedge\ $(($hv == ho$)$\ \implies\ t_{\gt{va}} == 0$)  & [0.902, 1] & 45.4 & 33.1 & 29.4\\\cline{3-8}
    \multirow{7}{*}{} & \multirow{5}{*}{}& EV & E[$\leqslant$3000; 500]([ ] $max:obs.t$)  & 667$\pm$79 & 80.8 & 29.7 & 31.7\\\cline{3-8}
    \multirow{7}{*}{} & \multirow{5}{*}{}& SI & simulate 500 [$\leqslant$3000]($hv, ho, v$))  & valid & 88.6 & 29.5 & 31.0\\\cline{2-8}
    \multirow{7}{*}{}& \multirow{2}{*}{R10}  &HT & Pr[$\leqslant$3000]([ ] $ha \leqslant ho\ \wedge\ $(($ha == ho$)$\ \implies\ t_{\gt{acc}} == 0$)$\geqslant$0.95  & valid & 2.4h & 44.7 & 29.2 \\\cline{3-8}
    \multirow{7}{*}{} & \multirow{2}{*}{} & PE & Pr[$\leqslant$3000]([ ] $ha \leqslant ho\ \wedge\ $(($ha == ho$)$\ \implies\ t_{\gt{acc}} == 0$)  & [0.902, 1] & 57.6 & 43.4 & 28.7 \\\cline{2-8}
    \multirow{7}{*}{}& \multirow{2}{*}{R11}  &HT & Pr[$\leqslant$3000]([ ] $htl \leqslant ho\ \wedge\ $(($htl == ho$)$\ \implies\ t_{\gt{tl}} == 0$)$\geqslant$0.95  & valid & 1.8h & 46.3 & 31.3 \\\cline{3-8}
    \multirow{7}{*}{} & \multirow{2}{*}{} & SI & simulate 100 [$\leqslant$3000]($htl, ho, tl$))  & valid & 56.2 & 42.4 & 30.7 \\\cline{2-8}
    \multirow{7}{*}{}& \multirow{2}{*}{R12}  &PE & Pr[$\leqslant$3000]([ ] $htr \leqslant ho\ \wedge\ $(($htr == ho$)$\ \implies\ t_{\gt{tr}} == 0$)  & [0.902, 1] & 52.9 & 44.1 & 31.0 \\\cline{3-8}
    \multirow{7}{*}{} & \multirow{2}{*}{} & SI & simulate 100 [$\leqslant$3000]($htr, ho, tr$))  & valid & 56.7 & 41.7 & 29.8 \\
   \hline
   \multirow{7}{*}{Synchronization} & \multirow{4}{*}{R13} & HT & Pr[$\leqslant$3000]([ ] $h_{dinf} \geqslant h_{sup}$) $\geqslant$ 0.95  & valid & 53.9 & 32.7 & 31.9\\\cline{3-8}
    \multirow{7}{*}{} & \multirow{5}{*}{}& PE & Pr[$\leqslant$3000]([ ] $h_{dinf} \geqslant h_{sup}$)  & [0.902, 1] & 13.7 & 35.5 & 30.4\\\cline{3-8}
    \multirow{7}{*}{} & \multirow{5}{*}{}& EV & E[$\leqslant$3000; 500]([ ] $max:checksync.t$)  & 30.6$\pm$0.21 & 72.4 & 32.6 & 31.6\\\cline{3-8}
    \multirow{7}{*}{} & \multirow{5}{*}{}& SI & simulate 500 [$\leqslant$3000]($h_{dinf}, h_{sup}$)  & valid & 86.8 & 32.6 & 32.0\\\cline{2-8}
    \multirow{7}{*}{}& \multirow{2}{*}{R14}  &PE & Pr[$\leqslant$3000]([ ] $h_{codinf} \geqslant h_{cosup}$)  & [0.902, 1] & 13.9 & 36.4 & 34.4 \\\cline{3-8}
    \multirow{7}{*}{} & \multirow{2}{*}{} & SI & simulate 100 [$\leqslant$3000]($h_{codinf}, h_{cosup}$)  & valid & 41.8 & 37.5 & 33.4 \\\cline{2-8}
    \multirow{7}{*}{}& \multirow{2}{*}{R15}  &PE & Pr[$\leqslant$3000]([ ] $h_{vidinf} \geqslant h_{visup}$)  & [0.902, 1] & 14.3 & 40.5 & 35.1 \\\cline{3-8}
    \multirow{7}{*}{} & \multirow{2}{*}{} & EV & E[$\leqslant$3000; 100]([ ] $max:checksyncvd.t$)  & 16.5$\pm$0.2 & 19.4 & 46.5 & 25.5 \\\cline{2-8}
    \multirow{7}{*}{}& \multirow{2}{*}{R16}  &HT & Pr[$\leqslant$3000]([ ] $h_{vddinf} \geqslant h_{vdsup}$)$\geqslant$0.95  & valid & 55.2 & 45.3 & 32.1 \\\cline{3-8}
    \multirow{7}{*}{} & \multirow{2}{*}{} & PE & Pr[$\leqslant$3000]([ ] $h_{vddinf} \geqslant h_{vdsup}$)  & [0.902, 1] & 13.9 & 40.7 & 33.5 \\
   \hline
    \end{tabular}%
  \label{table_verification_result}%
\end{table*}%

\begin{table*}[htbp]
 \scriptsize
  \centering
   \renewcommand\arraystretch{1}
\begin{tabular}[htbp]{|c|c|c|p{220pt}|c|c|c|c|}
    \hline
    Type & R.ID & Q & Expression & Result & Time & Mem & CPU \\
    \hline
   \multirow{7}{*}{End-to-End} & \multirow{5}{*}{R17} &HT  & Pr[$\leqslant$3000]$([\ ] h_{{\gt{lower}}} \geqslant h_{{\gt{spOut}}} \wedge ((h_{{\gt{lower}}}==h_{{\gt{spOut}}})\implies t_{{\gt{spOut}}}==0))$ $\geqslant$ 0.95  & valid & 54.2 & 32.9 & 31.4\\\cline{3-8}
    \multirow{7}{*}{} & \multirow{5}{*}{}& PE & Pr[$\leqslant$3000]$([\ ] h_{{\gt{lower}}} \geqslant h_{{\gt{spOut}}} \wedge ((h_{{\gt{lower}}}==h_{{\gt{spOut}}})\implies \neg t_{{\gt{spOut}}}))$  & [0.902, 1] & 13.1 & 35.3 & 29.4\\\cline{3-8}
    \multirow{7}{*}{} & \multirow{5}{*}{}& HT & Pr[$\leqslant$3000]$([\ ] h_{{\gt{spOut}}} \geqslant h_{{\gt{upper}}} \wedge ((h_{{\gt{spOut}}}==h_{{\gt{upper}}})\implies\ t_{{\gt{upper}}} ==0))$ $\geqslant$ 0.95  & valid & 1.3h & 32.2 & 32.6\\\cline{3-8}
    \multirow{7}{*}{} & \multirow{5}{*}{}& PE & Pr[$\leqslant$3000]$([\ ] h_{{\gt{spOut}}} \geqslant h_{{\gt{upper}}} \wedge ((h_{{\gt{spOut}}}==h_{{\gt{upper}}})\implies \neg t_{{\gt{upper}}}))$  & [0.902, 1] & 19.8 & 34.1 & 32.0\\\cline{3-8}
    \multirow{7}{*}{} & \multirow{5}{*}{}& EV & E[$\leqslant$3000; 500]([ ] $max:checke2e.t$)  & 229.7$\pm$0.9 & 83.3 & 32.5 & 30.6 \\\cline{3-8}
    \multirow{7}{*}{} & \multirow{5}{*}{}& SI & simulate 500 [$\leqslant$3000]($h_{{\gt{CU}}}, h_{{\gt{VD}}}, h_{{\gt{CL}}},t_{{\gt{CU}}},t_{{\gt{VD}}}$)  & valid & 89.8 & 32.9 & 30.2\\\cline{2-8}
    \multirow{7}{*}{}& \multirow{4}{*}{R18}  &HT & Pr[$\leqslant$3000]$([\ ] h_{{\gt{caml}}} \geqslant h_{{\gt{signOut}}} \wedge ((h_{{\gt{conl}}}==h_{{\gt{signOut}}})\implies t_{{\gt{signOut}}}==0))$ $\geqslant$ 0.95  & valid & 3.1h & 45.33 & 31.3 \\\cline{3-8}
    \multirow{7}{*}{} & \multirow{4}{*}{}& HT & Pr[$\leqslant$3000]$([\ ] h_{{\gt{camu}}} \leqslant h_{{\gt{signOut}}} \wedge ((h_{{\gt{conu}}}==h_{{\gt{signOut}}})\implies t_{{\gt{conu}}}==0))$ $\geqslant$ 0.95  & valid & 56.6 & 46.7 & 31.6 \\\cline{3-8}
    \multirow{7}{*}{} & \multirow{4}{*}{} & SI & simulate 100 [$\leqslant$3000]($h_{{\gt{camu}}}, h_{{\gt{signOut}}}, t_{{\gt{caml}}}$) & valid & 50.5 & 39.9 & 28.6 \\\cline{2-8}
    \multirow{7}{*}{}& \multirow{3}{*}{R19} & PE & Pr[$\leqslant$3000]$([\ ] h_{{\gt{caml}}} \geqslant h_{{\gt{vdOut}}} \wedge ((h_{{\gt{caml}}}==h_{{\gt{vdOut}}})\implies t_{{\gt{vdOut}}}==0))$  & [0.902, 1] & 52.7 & 39.3 & 30.4 \\\cline{3-8}
    \multirow{7}{*}{} & \multirow{3}{*}{} & PE & Pr[$\leqslant$3000]$([\ ] h_{{\gt{camu}}} \leqslant h_{{\gt{vdOut}}} \wedge ((h_{{\gt{camu}}}==h_{{\gt{vdOut}}})\implies t_{{\gt{camu}}}==0))$  & [0.902, 1] & 2.4h & 45.6 & 30.2 \\\cline{3-8}
    \multirow{7}{*}{} & \multirow{3}{*}{} & SI & simulate 100 [$\leqslant$3000]($h_{{\gt{camu}}}, h_{{\gt{vdOut}}}, t_{{\gt{caml}}}$)  & valid & 1.9h & 40.8 & 29.8 \\\cline{2-8}
    \multirow{7}{*}{}& \multirow{2}{*}{R20}  &HT & Pr[$\leqslant$3000]$([\ ] h_{{\gt{L}}} \geqslant h_{{\gt{tl}}} \wedge ((h_{{\gt{tl}}}==h_{{\gt{L}}})\implies t_{{\gt{tl}}}==0))$ $\geqslant$ 0.95  & valid & 151.3 & 41.9 & 29.1 \\\cline{3-8}
    \multirow{7}{*}{} & \multirow{2}{*}{} & SI & simulate 100 [$\leqslant$3000]($h_{{\gt{L}}}, h_{{\gt{tl}}}, t_{{\gt{tl}}}$)  & valid & 58.4 & 37.3 & 24.5 \\\cline{2-8}
    \multirow{7}{*}{}& \multirow{2}{*}{R21}  &HT & Pr[$\leqslant$3000]$([\ ] h_{{\gt{R}}} \geqslant h_{{\gt{tr}}} \wedge ((h_{{\gt{tl}}}==h_{{\gt{R}}})\implies t_{{\gt{tr}}}==0))$ $\geqslant$ 0.95  & valid & 75.9 & 46.8 & 31.3 \\\cline{3-8}
    \multirow{7}{*}{} & \multirow{2}{*}{} & SI & simulate 100 [$\leqslant$3000]($h_{{\gt{R}}}, h_{{\gt{tr}}}, t_{{\gt{tr}}}$)  & valid & 64.8 & 41.8 & 32.0 \\\cline{2-8}
    \multirow{7}{*}{}& \multirow{2}{*}{R22}  &PE & Pr[$\leqslant$3000]$([\ ] h_{{\gt{St}}} \geqslant h_{{\gt{st}}} \wedge ((h_{{\gt{St}}}==h_{{\gt{st}}})\implies t_{{\gt{st}}}==0))$  & [0.902, 1] & 18.5 & 41.9 & 27.3 \\\cline{3-8}
    \multirow{7}{*}{} & \multirow{2}{*}{} & SI & simulate 100 [$\leqslant$3000]($h_{{\gt{St}}}, h_{{\gt{st}}}, t_{{\gt{st}}}$)  & valid & 57.5 & 36.9 & 33.5 \\\cline{2-8}
    \multirow{7}{*}{}& \multirow{2}{*}{R23}  &PE & Pr[$\leqslant$3000]$([\ ] h_{{\gt{Stop}}} \geqslant h_{{\gt{stu}}} \wedge ((h_{{\gt{Stop}}}==h_{{\gt{stu}}})\implies t_{{\gt{stu}}}==0))$  & [0.902, 1] & 26.8 & 42.3 & 27.8 \\\cline{3-8}
    \multirow{7}{*}{} & \multirow{2}{*}{} & SI & simulate 100 [$\leqslant$3000]($h_{{\gt{Stop}}}, h_{{\gt{stu}}}, t_{{\gt{stu}}}$)  & valid & 73.6 & 42.4 & 27.9 \\
   \hline
   \multirow{7}{*}{Comparison} & \multirow{5}{*}{R24} &HT & Pr[$\leqslant$3000]([ ] ($ex_{{\gt{con}}} == wcet_{{\gt{con}}}\ \wedge\ ex_{{\gt{vd}}} == wcet_{{\gt{vd}}}$)$\ \implies\ $($h_{{\gt{cu}}} \geqslant h_{{\gt{com}}}$)) $\geqslant$ 0.95  & valid & 57.4 & 36.7 & 28.4\\\cline{3-8}
    \multirow{7}{*}{} & \multirow{5}{*}{}& PE & Pr[$\leqslant$3000]([ ] ($ex_{{\gt{con}}} == wcet_{{\gt{con}}}\ \wedge\ ex_{{\gt{vd}}} == wcet_{{\gt{vd}}}$)$\ \implies\ $($h_{{\gt{cu}}} \geqslant h_{{\gt{com}}}$))  & [0.902, 1] & 14.7 & 35.5 & 26.7\\\cline{3-8}
    \multirow{7}{*}{} & \multirow{5}{*}{}& EV & E[$\leqslant$3000; 500]([ ] $max:control.t$)  & 146.7$\pm$0.28 & 74.9 & 29.4 & 32.7\\\cline{3-8}
    \multirow{7}{*}{} & \multirow{5}{*}{}& EV & E[$\leqslant$3000; 500]([ ] $max:vd.t$)  & 96.6$\pm$0.27 & 74.2 & 29.4 & 31.4\\\cline{3-8}
    \multirow{7}{*}{} & \multirow{5}{*}{}& SI & simulate 500 [$\leqslant$3000]($h_{\gt{cu}}, h_{\gt{com}}$))  & valid & 86.6 & 29.5 & 32.5\\\cline{2-8}
    \multirow{7}{*}{}& \multirow{2}{*}{R25}  &EV & E[$\leqslant$3000; 100]([ ] $max:camera.t$)  & 29.8$\pm$0.02 & 18.7 & 39.6 & 29.9 \\\cline{3-8}
    \multirow{7}{*}{} & \multirow{2}{*}{} & EV & E[$\leqslant$3000; 100]([ ] $max:signreg.t$)  & 143.5$\pm$0.7 & 16.5 & 33.2 & 28.7 \\\cline{3-8}
    \multirow{7}{*}{} & \multirow{2}{*}{} & SI & simulate 100 [$\leqslant$3000]($h_{\gt{sigu}}, h_{\gt{sig}}$))  & valid & 12.6 & 35.6 & 29.8 \\\cline{2-8}
    \multirow{7}{*}{}& \multirow{2}{*}{R26}  &HT & Pr[$\leqslant$3000]([ ] $ex_{{\gt{con}}} == wcet_{{\gt{con}}}\ \wedge\ ex_{{\gt{vd}}} == wcet_{{\gt{vd}}}\ \wedge\ ex_{{\gt{cam}}} == wcet_{{\gt{cam}}}\ \wedge\ ex_{{\gt{sign}}} == wcet_{{\gt{sign}}}$)$\ \implies\ $($h_{{\gt{au}}} \geqslant h_{{\gt{a}}}$)$\geqslant$0.95  & valid & 2.1h & 42.5 & 30.1 \\\cline{3-8}
    \multirow{7}{*}{} & \multirow{2}{*}{} & PE & Pr[$\leqslant$3000]([ ] $ex_{{\gt{con}}} == wcet_{{\gt{con}}}\ \wedge\ ex_{{\gt{vd}}} == wcet_{{\gt{vd}}}\ \wedge\ ex_{{\gt{cam}}} == wcet_{{\gt{cam}}}\ \wedge\ ex_{{\gt{sign}}} == wcet_{{\gt{sign}}}$)$\ \implies\ $($h_{{\gt{au}}} \geqslant h_{{\gt{a}}}$)  & [0.902, 1] & 56.9 & 40.7 & 29.7 \\
   \hline
   \multirow{7}{*}{Exclusion} & \multirow{4}{*}{R27} &HT  & Pr[$\leqslant$3000]([ ] $\neg (t_{\gt{Right}} == 1\ \wedge\ t_{\gt{Left}} == 1)$) $\geqslant$ 0.95  & valid & 57.4 & 36.7 & 28.4\\\cline{3-8}
    \multirow{7}{*}{} & \multirow{4}{*}{} & PE & Pr[$\leqslant$3000]([ ] $\neg (t_{\gt{Right}} == 1\ \wedge\ t_{\gt{Left}} == 1)$)  & [0.902, 1] & 14.7 & 35.5 & 26.7\\\cline{3-8}
    \multirow{7}{*}{} & \multirow{4}{*}{} & PC & Pr[$\leqslant$3000]([ ] $\neg (t_{\gt{Right}} == 1\ \wedge\ t_{\gt{Left}} == 1)$) $\geqslant$ Pr[$\leqslant$3000]($<> \neg$($\neg (t_{\gt{Right}} == 1\ \wedge\ t_{\gt{Left}} == 1)$))  & $\geqslant$1.1 & 10.9 & 34.2 & 31.3\\\cline{3-8}
    \multirow{7}{*}{} & \multirow{4}{*}{} & SI & simulate 500 [$\leqslant$3000]($t_{\gt{Right}}, t_{\gt{Left}}$)  & valid & 85.5 & 29.6 & 32.6\\\cline{2-8}
    \multirow{7}{*}{}& \multirow{2}{*}{R28}  &HT & Pr[$\leqslant$3000]([ ] $\neg (t_{\gt{b}} == 1\ \wedge\ t_{\gt{acc}} == 1)$)$\geqslant$0.95  & valid & 57.5 & 44.6 & 35.9 \\\cline{3-8}
    \multirow{7}{*}{} & \multirow{2}{*}{} & PE & Pr[$\leqslant$3000]([ ] $\neg (t_{\gt{b}} == 1\ \wedge\ t_{\gt{acc}} == 1)$)  & [0.902, 1] & 14.3 & 40.7 & 35.6 \\\cline{2-8}
    \multirow{7}{*}{}& \multirow{2}{*}{R29}  &HT & Pr[$\leqslant$3000]([ ] $\neg (t_{\gt{eme}} == 1\ \wedge\ t_{\gt{Left}} == 1)$)$\geqslant$0.95  & valid & 62.6 & 40.6 & 33.6 \\\cline{3-8}
    \multirow{7}{*}{} & \multirow{2}{*}{} & SI & simulate 100 [$\leqslant$3000]($t_{\gt{eme}}, t_{\gt{Left}}$)  & valid & 46.5 & 36.4 & 34.2 \\\cline{2-8}
    \multirow{7}{*}{}& \multirow{2}{*}{R30}  &HT & Pr[$\leqslant$3000]([ ] $\neg (t_{\gt{eme}} == 1\ \wedge\ t_{\gt{Right}} == 1)$)$\geqslant$0.95  & valid & 63.8 & 36.3 & 34.2 \\\cline{3-8}
    \multirow{7}{*}{} & \multirow{2}{*}{} & SI & simulate 100 [$\leqslant$3000]($t_{\gt{Right}}, t_{\gt{eme}}$)  & valid & 47.7 & 36.4 & 34.5 \\\cline{2-8}
    \multirow{7}{*}{}& \multirow{2}{*}{R31}  &HT & Pr[$\leqslant$3000]([ ] $\neg (t_{\gt{eme}} == 1\ \wedge\ t_{\gt{acc}} == 1)$)$\geqslant$0.95  & valid & 59.1 & 36.3 & 35.2 \\\cline{3-8}
    \multirow{7}{*}{} & \multirow{2}{*}{} & PE & Pr[$\leqslant$3000]([ ] $\neg (t_{\gt{eme}} == 1\ \wedge\ t_{\gt{acc}} == 1)$)  & [0.902, 1] & 15.5 & 36.7 & 30.1 \\
   \hline

    \end{tabular}%
\end{table*}%
We have formally analyzed over 30 properties (associated with timing constraints) of the system including \emph{deadlock freedom}. A list of selected properties (Chapter 3) are verified using \smc\ and the results are listed in Table.\ref{table_verification_result}.
Five types of \smc\ queries are employed to specify R1 -- R31, \emph{Hypothesis Testing} (HT), \emph{Probability Estimation} (PE), \emph{Probability Comparison} (PC), \emph{Expected Value} (EV) and \emph{Simulations} (SI).

\begin{figure}[htbp]
\centerline{{\includegraphics[width=3.3in]{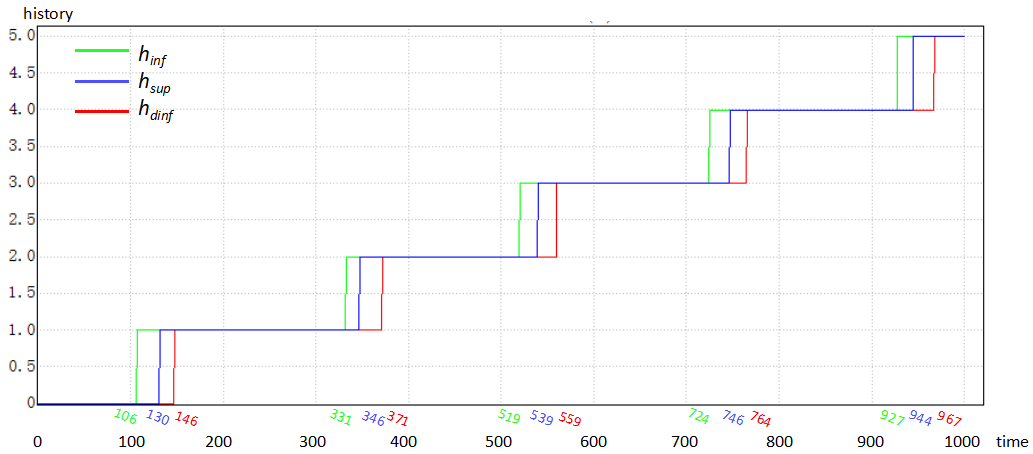}
}}
\caption{Simulation Result of R13}
\label{fig:simulation}
\hfil
\end{figure}

\begin{figure}[htbp]
\centerline{{\includegraphics[width=3.3in]{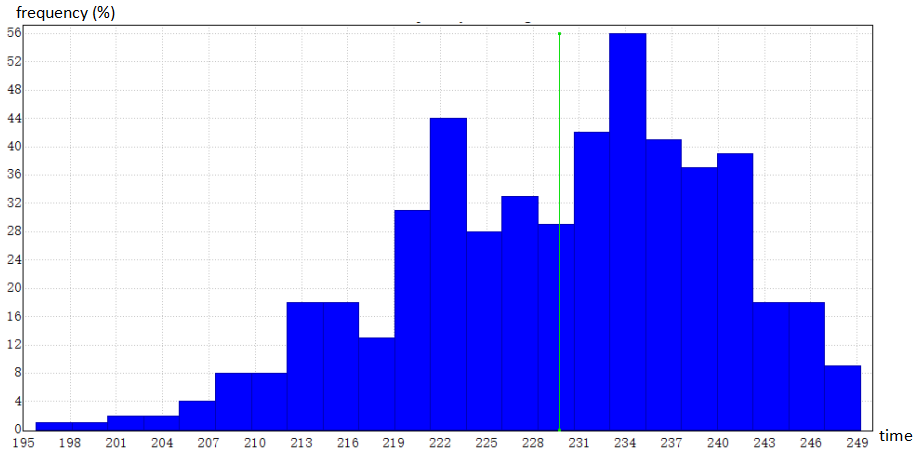}
}}
\caption{Frequency Histogram of {\gt{End-to-End}} timing constraint (R17)}
\label{fig:EV}
\hfil
\end{figure}

\begin{inparaenum}
\item \emph{Deadlock Freedom}: Because of the insufficient memory caused by the periodically triggered STA {\gt{ms}}, the \emph{Deadlock Freedom} property cannot be checked successfully.
\item \emph{Hypothesis Testing}: All properties are established as valid with 95\% level of confidence; \item  \emph{Probability Estimation}: The probability of each property being satisfied is computed and its approximate interval is given as [0.902, 1]; \item \emph{Expected Value}: The expected values of time durations of timing constraints (R1, R2, R5, R9, R13, R15, R17, R24 -- R25) are evaluated.
    For example, during the analysis of R1, the time interval between two consecutive triggerings of the {\gt{Camera}} is evaluated as 50 and that validates R1. Furthermore, \smc\ evaluates the expected maximum duration bound of {\gt{End-to-End}} timing constraint by checking R17 and generates the frequency histogram of the expected bound (see Fig. \ref{fig:EV}). It illustrates that the expected bound is always less than 250ms and 90\% of the duration is within the range of [207, 249]; \item \emph{Probability Comparison}: is applied to confirm that the probability of {\gt{SignRecognition}} \fp\ completing its execution within [100, 125]ms is greater than the probability of completion within [125, 150]ms (R5). The query results in a comparison probability ratio greater than or equal to 1.1, i.e., the execution time of \gt{SignRecognition} \fp\ is most likely less than 125ms. Similarly, R6 -- R8 can be analyzed. \item \emph{Simulation}: The simulation result of {\gt{Synchronization}} timing constraint (R13) is demonstrated in Fig. \ref{fig:simulation}. $h_{inf}$, $h_{sup}$ and $h_{dinf}$ are history of $inf$, $sup$ and $dinf$ respectively. Recall Spec. R13 (see Fig. \ref{fig:AV_model}), the {\gt{causality}} \emph{relation} between $dinf$ and $sup$ is satisfied. As the simulation of R13 shows (Fig. \ref{fig:simulation}), the rising edge of $h_{sup}$ (in blue) always occurs prior to $h_{dinf}$ (in red). It indicates that $sup$ always runs faster than $dinf$, thus the \gt{causality} \emph{relation} is validated.
\end{inparaenum}

\begin{figure}[htbp]
\centering
  \subfigure[\emph{Simulation}]{
  \includegraphics[width=2.5in]{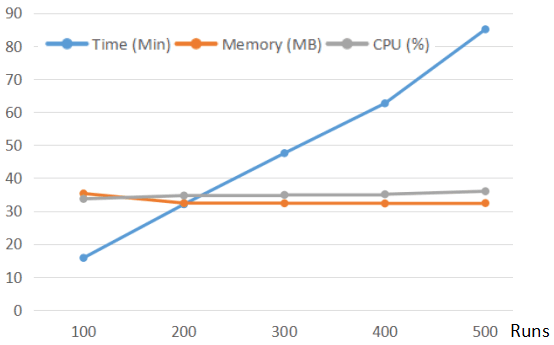}
  \label{fig:simu}}
  \subfigure[\emph{Expected Value}]{
  \includegraphics[width=2.5in]{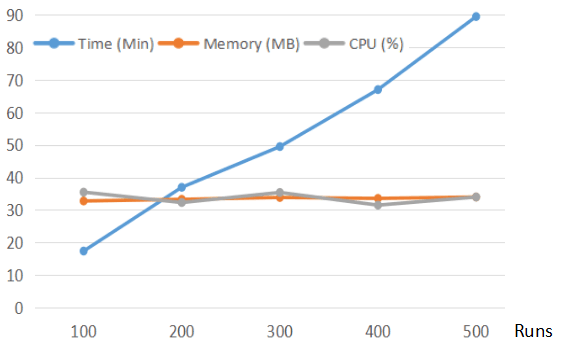}
  \label{fig:ev}}
  \caption{Performance analysis of verifying R5 with \emph{Expected Value} and \emph{Simulation}. The number of runs ranges from 100 to 500 with increment as 100.}
\label{fig:performance}
\end{figure}

We estimate the performance (i.e., time, memory and CPU consumption) of verifying R5 by using \emph{Expected Value} and \emph{Simulation} queries with different numbers of runs assigned. As shown in Fig. \ref{fig:performance}, along with the increase of the number of runs, for both queries, the verification time grows proportionally, while the CPU and memory have no significant changes.

\chapter{Related work}
\label{sec:r-work}

In the context of \ed, efforts on the integration of \ed\ and formal techniques based on timing constraints were investigated in several works  \cite{kress13,qureshi2011,ksafecomp11,Goknil2013Analysis}, which are however, limited to the executional aspects of system functions without addressing stochastic behaviors. Kang \cite{ksac14} and Suryadevara \cite{Suryadevara2013Validating, Suryadevara2013Verifying} defined the execution semantics of both the controller and the environment of industrial systems in \ccsl\ which are also given as mapping to \uppaal\ models amenable to model checking. In contrast to our current work, those approaches lack precise stochastic annotations specifying continuous dynamics in particular regarding different clock rates during execution. Ling \cite{Yin2011Verification} transformed a subset of \ccsl\ constraints to PROMELA models to perform formal verification using SPIN. Zhang \cite{Zhang2017Towards} transformed \ccsl\ into first order logics that are verifiable using SMT solver. However, their works are limited to functional properties, and no timing constraints are addressed. Though, Kang et al. \cite{kiciea16,kapsec15} and Marinescu et al. \cite{Marinescu3762} presented both simulation and model checking approaches of \simu\ and \smc\ on \ed\ models, neither formal specification nor verification of extended \ed\ timing constraints with probability were conducted. Our approach is a first application on the integration of \ed\ and formal V\&V techniques based on  probabilistic extension of \ed/\tdl\ constraints using Pr\ccsl\ and \smc.
An earlier study \cite{mvv, sscps, sac18}
defined a probabilistic extension of \ed\ timing constraints and presented model checking approaches on \ed\ models, which inspires our current work. Specifically, the techniques provided in this paper define new operators of \ccsl\ with stochastic extensions (Pr\ccsl) and verify the extended \ed\ timing constraints of CPS (specified in Pr\ccsl) with statistical model checking.
Du. et al. \cite{Du2016MARTE} proposed the use of \ccsl\ with probabilistic logical clocks to enable stochastic analysis of hybrid systems by limiting the possible solutions of clock ticks. Whereas, our work is based on the probabilistic extension of \ed\ timing constraints with a focus on probabilistic verification of the extended constraints, particularly, in the context of WH.

\chapter{Conclusion}
\label{sec:conclusion}

We present an approach to perform probabilistic verification on \ed\ timing constraints of automotive systems based on WH at the early design phase: \begin{inparaenum} \item Probabilistic extension of \ccsl, called Pr\ccsl, is
defined and the \ed/\tdl\ timing constraints with stochastic properties
are specified in Pr\ccsl; \item The semantics of the extended constraints in Pr\ccsl\ is translated into verifiable \smc\ models for formal verification; \item A set of mapping rules is
proposed to facilitate guarantee of translation. \end{inparaenum} Our approach is demonstrated on an autonomous traffic sign recognition vehicle (AV) case study.
Although, we have shown that defining and translating a subset of \ccsl\ with probabilistic extension into \smc\ models is sufficient to verify \ed\ timing constraints, as ongoing work, advanced techniques covering a full set of \ccsl\ constraints are further studied.
Despite the fact that \smc\ supports probabilistic analysis of the timing constraints of AV, the computational cost of verification in terms of time is rather expensive. Thus, we continue to investigate complexity-reducing design/mapping patterns for CPS to improve effectiveness and scalability of system design and verification.

\chapter*{Acknowledgment}
This work is supported by the NSFC, EASY Project: 46000-41030005.

\end{document}